\documentclass[preprint]{aastex63}
\usepackage{amsmath}
\usepackage{graphicx}
\usepackage{tikz}
\usepackage{booktabs}
\usepackage{multimedia}

\received{??, 2020}
\revised{??, 2020}
\accepted{??, 2020}
\submitjournal{}
\shorttitle{Brightening Detection} 
\shortauthors{Humphries, Morgan}
\setwatermarkfontsize{240pt}
\watermark{DRAFT}

\newcommand{\de}{$^{\circ}$}

\newcommand{\kms}{~km\,$s^{-1}$}

\newcommand{\paperi}{Paper I}

\newcommand{\nfrag}{$N_{frag}$}
\newcommand{\threshlo}{$T_{low}$}
\newcommand{\threshhi}{$T_{high}$}

\begin{document}

\title{A multi-wavelength analysis of small-scale brightenings observed by IRIS }
\correspondingauthor{Humphries,  Morgan}
\email{llh18@aber.ac.uk, hum2@aber.ac.uk}

\author[0000-0002-0786-7307]{Ll\^yr Dafydd Humphries}
\affiliation{Aberystwyth University \\
Faculty of Business and Physical Sciences\\
Aberystwyth, Ceredigion, SY23 3FL, Wales, UK}

\author[0000-0002-0786-7307]{Huw Morgan}
\affiliation{Aberystwyth University \\
Faculty of Business and Physical Sciences\\
Aberystwyth, Ceredigion, SY23 3FL, Wales, UK}

\begin{abstract}
Small-scale brightenings in solar atmospheric observations are a manifestation of heating and/or energy transport events. We present statistical characteristics of brightenings from a new detection method applied to 1330, 1400, and 2796 \AA\ IRIS slitjaw image time series. 2377 events are recorded which coexist in all three channels, giving high confidence that they are real. $\approx$1800 of these are spatially coherent, equating to event densities of $\sim9.7\times10^{-5}$arcsec$^{-2}$s$^{-1}$ within a $90\arcsec\times100\arcsec$ FOV over 34.5 minutes. Power Law indices estimates are determined for total brightness ($2.78<\alpha<3.71$), maximum brightness ($3.84<\alpha<4.70$), and average area ($4.31<\alpha<5.70$) distributions. Duration and speed distributions do not obey a power law. A correlation is found between the events' spatial fragmentation, area, and duration, and a weak relationship with total brightness, showing that larger/longer-lasting events are more likely to fragment during their lifetime. Speed distributions show that all events are in motion, with an average speed of $\sim7$\kms. The events' spatial trajectories suggest that cooler 2796 \AA\ events tend to appear slightly later, and occupy a different position/trajectory to the hotter channel results. This suggests that either many of these are impulsive events caused by reconnection, with subsequent rapid cooling, or that the triggering event occurs near the TR, with a subsequent propagating disturbance to cooler atmospheric layers. The spatial distribution of events is not uniform, with broad regions devoid of events. A comparison of spatial distribution with properties of other atmospheric layers shows a tentative connection between high magnetic field strength, the corona's multithermality, and high IRIS brightening activity.

\end{abstract}

%======================
% Introduction
%======================

\section{Introduction}
\label{sec:study_3_intro}
Despite the wealth of imaging and spectral observations of the lower solar atmosphere, comprehensive, homogeneous, multi-wavelength studies of small-scale structures are relatively rare. This is partly due to the difficulty of automatically detecting these events: the largest number of events will be at the resolution limit and noise level of observations, offering a substantial challenge to detection methods. Some examples of recent studies include \cite{hou}'s spectral analysis of over 2700 AR \ion{Si}{4} dots, \cite{Parnell_2000}'s TRACE study of over 20,000 nano-flare-like events, and \cite{Hu_2018}'s observations of $\sim74,000$ small-scale flux ropes over two solar cycles. Several studies of sub-arsecond bright dots and small-scale jets are near to the current resolution limit of solar instruments \citep{Huang_2014, Tian_2014, Tiwari_2019, Huang_2021}. 

Numerous Ellerman Bomb (EB) studies have been conducted \citep[][,etc.]{McMath_1960, vissers_2015, Chen_2019}, with \cite{nelson_2013} an example of a large-scale study of 3570 H$\alpha$ observations. The driving mechanism of these small (of the order $1\arcsec$), and short-lived \citep[$\sim20$ minutes,][]{Severny_1964} photospheric brightenings is greatly debated. \cite{Stellmacher_1991} attribute the asymmetric spectral results of EBs to a super-position of a hot plasma component surrounded by a cooler intergranular region, such as in the vicinity of the penumbra of isolated sunspots. \cite{Kitai_1984}'s view is that EBs occur in funnel-like, individual flux tubes, rather than as a result of magnetic reconnection. However, \cite{Georgoulis_2002}'s study suggests that stochastic magnetic reconnection may be the cause due to the turbulent nature of photospheric magnetic field lines. This is in spite of \cite{Pikelner_1974}'s argument that EBs should locally reach coronal temperatures following field-line annihilation while EBs typically demonstrate temperatures $<7000$K \citep{Gonzalez_2013}. 
Empirical studies such as these are complemented by numerical modelling and MHD simulation studies \citep[e.g.][]{Guerreiro_2015, Guerreiro_2017}. 

Similar phenomena present themselves in far ultraviolet (FUV) IRIS observations, namely IRIS bombs (IB). These small pockets of plasma are of similar size to EBs (a few arcseconds wide) and are also likely the result of magnetic reconnection and undulating field line submergence \citep{Peter_2014}. Both emission and absorption lines are present which are likely the result of these magnetic field submergences and plasma outbursts, respectively. However, these chromospheric bombs are far hotter than typical EB temperatures: \cite{Tian_2016} demonstrate temperatures of $8\times10^{4}$K. Some tenuous correlation between EBs and IBs is suggested although this remains mostly unclear.
The main challenge, given the current limit of observations, is to provide unambiguous interpretation of detected events, that is, what is their true cause and physical nature. The goal of this work is to detect and present general characteristics of a large number of brightening events in the chromosphere/transition region (TR), and we defer the more difficult task of classifying the events into different possible types for future work.

Many solar phenomena follow power-law distributions \citep{Aschwanden_2002, Aschwanden_2015, nhalil_2020}, large flares being the most commonly analysed \citep[e.g.][etc.]{Verbeeck_2019, Wheatland_2004, Li_2016}. $\alpha$ is a parameter that measures the exponent of a power law fit to an observed distribution. In order to provide sufficient energy to heat the corona, nano-flares, micro-flares or other small-scale phenomena must be numerous at small spatiotemporal scales that are below the resolution of current observations. $\alpha<2$ suggests that flare input power is dominated by the high-energy part of the thermal energy and large-scale events, and that small-scale events provide insufficient energy density to fully account for the solar atmosphere's heating, while the reverse is true for $\alpha>2$ \citep{Hudson_1991}. Typical examples of soft X-ray flare flux distributions lie within $1.75<\alpha<1.84$ \citep{Hudson_1969, Drake_1971}. Other peak flux observations provide power-law indices of $\sim1.8$ \citep{Datlowe_1974, Crosby_1993, Clauset_2007}. Some estimates of small-scale event indices reach as high as 3.5 for small flare prediction \citep{Vlahos_1995}. While recent studies challenge the ambiguities that accompany small-scale event analyses, their role in coronal heating are supported by numerous observations and numerical simulations \citep{Testa_2014, bowness_2013, Peter_2006} suggesting that small-scale heating scenarios are capable of generating a million-degree hot corona.

The work carried out in \citet{Humphries2021} (hereafter \paperi) demonstrated, using tests on synthetic and real IRIS quiet-sun (QS) data, that small-scale brightening events can be filtered from their background, detected, and their characteristics extracted with useful precision. Namely, compared to their `true' synthetic attributes, the statistical mean of their centroid positions, duration, maximum brightness and average speed were accurately replicated albeit with large uncertainties.
The total brightness and average area of each event was consistently underestimated. This is due to the nature of the thresholding method used in \paperi, and is likely true for most detection methods. Determining whether two close events are separate, related, or a defective detection of a single event, is important. It can be addressed with several approaches, an example being \cite{sekse}'s frame-by-frame pixel density overlap to determine whether events in neighbouring frames belong to the same chain of events. Determining whether two events overlap across two different datasets (e.g. two channels from the same instrument), is also challenging, as \cite{Henriques_david} demonstrated using a null hypothesis, Bernoulli distribution, and Chernoff bounds for Poisson trials. A statistical basis for whether two events are in fact the same event observed across multiple channels is presented in this paper based on the likelihood of the centroid of two random points overlapping within an area comparable to that being observed.

The goal of this study is to apply the brightening detection and characterising method from \paperi\ to an IRIS image time series in three wavelength channels. We note that the dataset used in this current study is different to the single-wavelength data used in \paperi. This study is concerned with presenting statistical aspects of all detected small-scale events in the dataset, and no effort is made to categorize detections into different types of events, or to interpret the cause or nature of the detections. The wealth of information returned on the thousands of detections can, in principle, give insight into the nature of the events, and this will require additional future work. The characteristics of these events across each wavelength are compared along with an investigation of the statistical likelihood of whether two events detected in different wavelengths represent the same event (using the event-by-event comparison criterion from \paperi). We also apply a statistical method for quantifying the power-law behaviour of these events. Section \ref{sec:obs} provides a brief summary of the observations and region of interest (ROI), and section \ref{sec:study_3_detection_method} gives an overview of the method and various parameters. Section \ref{sec:match} shows how events are paired across all three channels, and section \ref{sec:power} describes and discusses our choice of method for fitting the power law indices. Results are presented and discussed in section \ref{sec:study_3_results}, and conclusions are in section \ref{sec:study_3_conclusions}.

%------------------------------------------- 
%Method section 
%-------------------------------------------

\section{Observations \& Method}\label{sec:study_3_observations}

\subsection{Observations}
\label{sec:obs}

\begin{figure*}[t]
    \centering
    \includegraphics[trim={1cm 0cm 0cm 0cm},clip,width=\textwidth]{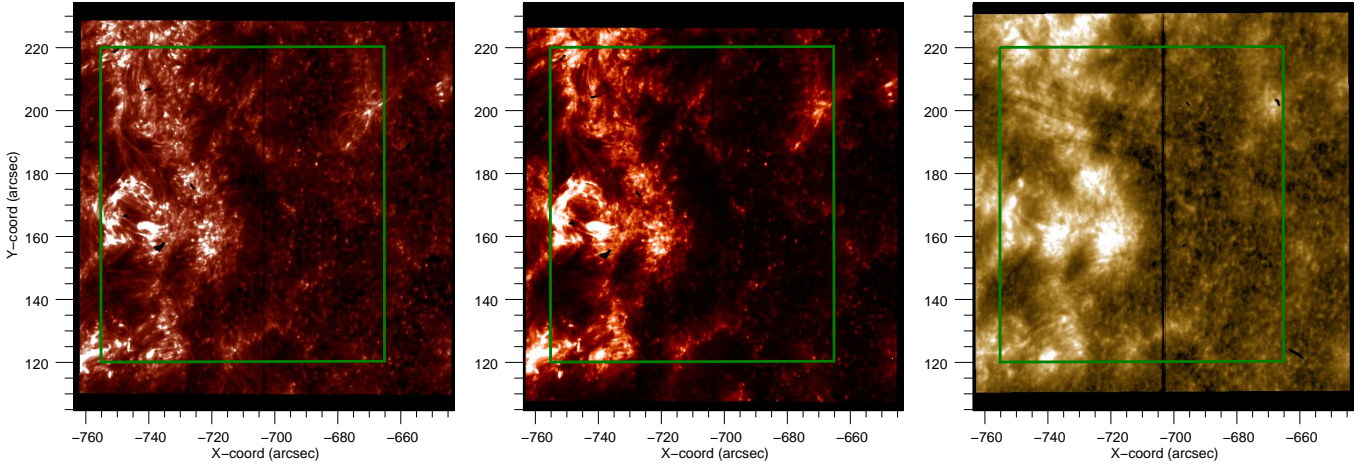}
    \caption{IRIS (left) 1400 \AA, (center) 1330 \AA\ and 2796 \AA\ slit-jaw images of the ROI on Dec. 18th 2014 at approximately 12:32 UT. Green box denotes the smaller FOV common to each wavelength channel. Each image has been processed such that isolated pixels or small regions of outlying intensity (such as the central slit) are treated as missing data.}
    \label{fig:fov}
\end{figure*}

We select an IRIS \citep{pontieu} dataset consisting of C {\sc{ii}} 1330 \AA\, Si {\sc{iv}} 1400 \AA\ and Mg {\sc{{ii}}} h/k 2796 \AA\ sit-and-stare series of slit-jaw images, each with a consistent 17s temporal cadence and 0.17$\arcsec \times 0.17\arcsec$ per pixel spatial scale. Figure \ref{fig:fov} shows the ROI centered on a small mid-latitude coronal hole within the North-East quadrant. The coronal hole is bounded by a small active region to the east. Some discrepancy is present between each channel's field of view (FOV) therefore a new, smaller ROI is chosen that is common for each wavelength, denoted by the green square in figure \ref{fig:fov}. This new ROI is centred at X = $-710\arcsec$, Y = $170\arcsec$ with a $90\arcsec\times100\arcsec$ FOV. 
The observations begins on Dec. 18th 2014 at 12:32UT and ends the same day at 13:05UT. 

The IRIS data set is processed using standard procedures to account for temporal exposure, and empty margins are removed. 
Dark current subtraction as well as flat field, geometrical and orbital variation corrections have been applied as standard procedure for level 2 IRIS data. Isolated pixels with spuriously high values are flagged as missing data, and are ignored in the filtering and detection method. The central slit and its surrounding pixels are also treated as missing data such that extremely `dark' pixels do not affect the filtering process. Additionally, small drifts due to solar rotation have been corrected using a Fourier Local Correlation Tracking method \citep{fw}.

\subsection{Detection Method}\label{sec:study_3_detection_method}

The filtering and thresholding procedure is the same as that used in \paperi. The 1400 \AA\ data cube first undergoes band-pass filtering in space and time by applying three sequential convolutions with vector kernels in the two spatial and one temporal dimension.
The filtered cube $F$ is defined as 
\begin{equation}
\begin{aligned}
S^\prime &=S \ast\alpha (x)\\
S^{\prime\prime} &=S^\prime \ast\beta (y)\\
F & = S^{\prime\prime} \ast\gamma (t) ,\\
\end{aligned}
\end{equation}
where $\alpha (x)$, $\beta (y)$ and $\gamma (t)$ are the $x$, $y$ and $t$ kernels respectively, as set by the Interactive Data Language (IDL) \textit{Digital\_filter} procedure \citep{Walraven}.
%The $x$ and $y$ kernels in this work are identical as the FOV is not located at the limb and therefore avoids the complication of differential line-of-sight.
The bandpass frequency range is defined by the $f_{low}$ and $f_{high}$ parameters, whereby frequencies above and below a particular frequency are attenuated such that the power of spatial/temporal slow-changing features (low frequency) and background noise (high frequency) are significantly reduced. The work in \paperi\ finds that $f_{low}=0.09$ and $f_{high}=0.40$ are optimal and all filtering conducted in this work uses these frequency limits.

Subsequently, a spatially-varying threshold is defined as a multiple of the standard deviation over time of a filtered analogue synthetic noise cube based on Poisson statistics (see \paperi\ for details on generating synthetic data cubes). This threshold is defined as \threshhi\ and any pixels above this threshold are considered as initial detection candidates.

Finally, a second, lower threshold, \threshlo\ is defined and applied to an extended region surrounding each initial detection. All voxels within the surrounding regions that are above \threshlo\ are considered part of the event detected above \threshhi, thus the initially small detected regions are grown. This enables the isolation of separate events, thus help avoid overlapping detections, and a more accurate reading of an event's area and total brightness. \threshlo=7 and \threshhi=9 were found to be optimal values for the 1400 \AA\ data in \paperi\ and are therefore used for the 1400 \AA\ data in this study. Due to intrinsic differences in the effective area of each slit-jaw channel (see \cite{pontieu} for details), the \threshhi\ and \threshlo\ values are adjusted for the 1330 \AA\ and 2976 \AA\
data sets. Using the 1400 \AA\ detections as a reference, the thresholds are adjusted such that the number of 1330 \AA\ and 2796 \AA\ detections match the number of 1400 \AA\ detections plus 10\%. These new threshold values and the corresponding number of detections are listed in table \ref{tab:study_3_thresh}, with \threshlo$=4.05$ and \threshhi$=5.13$ for 1330 \AA\, and \threshlo$=1.86$, \threshhi$=2.15$ for 2796 \AA. 

Several characteristics for each detected region are recorded including duration, average area, maximum brightness, total brightness, centroid position as a function of time, and average speed. An additional parameter is also recorded which describes the complexity of an event and whether that event fragments during its lifetime, denoted as $N_{frag}$ (see \paperi\ for further details). The whole spatiotemporal volume defining the event is always connected by one or more neighbouring voxels, however, during a single time step the event may occupy one or more isolated regions. $N_{frag}$ records the greatest number of isolated regions recorded during an event's lifetime. Candidate brightenings that are smaller than 25 voxels, and/or last less than 5 frames, are discarded. There are no upper limits to area/volume or duration provided that intensity changes differ enough from the background. There is an intrinsic upper limit to an event's motion: if an event is moving so rapidly that there is no spatial overlap between one  frame and another, the moving regions will not be labelled as the same event, and the individual regions will fail the minimum lifetime threshold and not be recorded, although this is highly unlikely given the very high speeds required to prevent overlap of the extended event area. Finally, intensity measurements are limited only by their significance with regards to the spatially varying threshold i.e. if a brightening is not filtered out as noise or slow background variations, then it is detected as a candidate event. Therefore, there are no direct rigid lower or upper intensity limits as long as the intensity is above the locally-defined threshold based on Poisson noise.

\subsection{Matching events across channels}
\label{sec:match}

Setting the detection thresholds so that the number of 1330 \AA\ and 2796 \AA\ candidate events are higher than the number of 1400 \AA\ detections produces false positives. However, this is greatly minimized by implementing an event-by-event distance criterion. \paperi\ used the spatiotemporal distance of detection centroids from the known synthetic brightenings, defined as $\Delta=\sqrt{\Delta x^2+\Delta y^2+\Delta t^2}$ (measured in units of pixels or time frame number), in order to test the method's detection accuracy. Here, we determine the minimum spatio-temporal separation between detections across multiple wavelengths, discard those with poor agreement, and thus reduce the number of 1330 \AA\ and 2796 \AA\ detections to directly match that of 1400 \AA. This greatly reduces the chance of false positives. %However, tests on a simple example of a synthetic cube test in \paperi\ suggest that $\sim0.7\%$ of all detections are false-positives. 

\begin{figure}
    \centering
    \includegraphics[trim={0cm 0cm 0cm 0cm},clip,width=\textwidth]{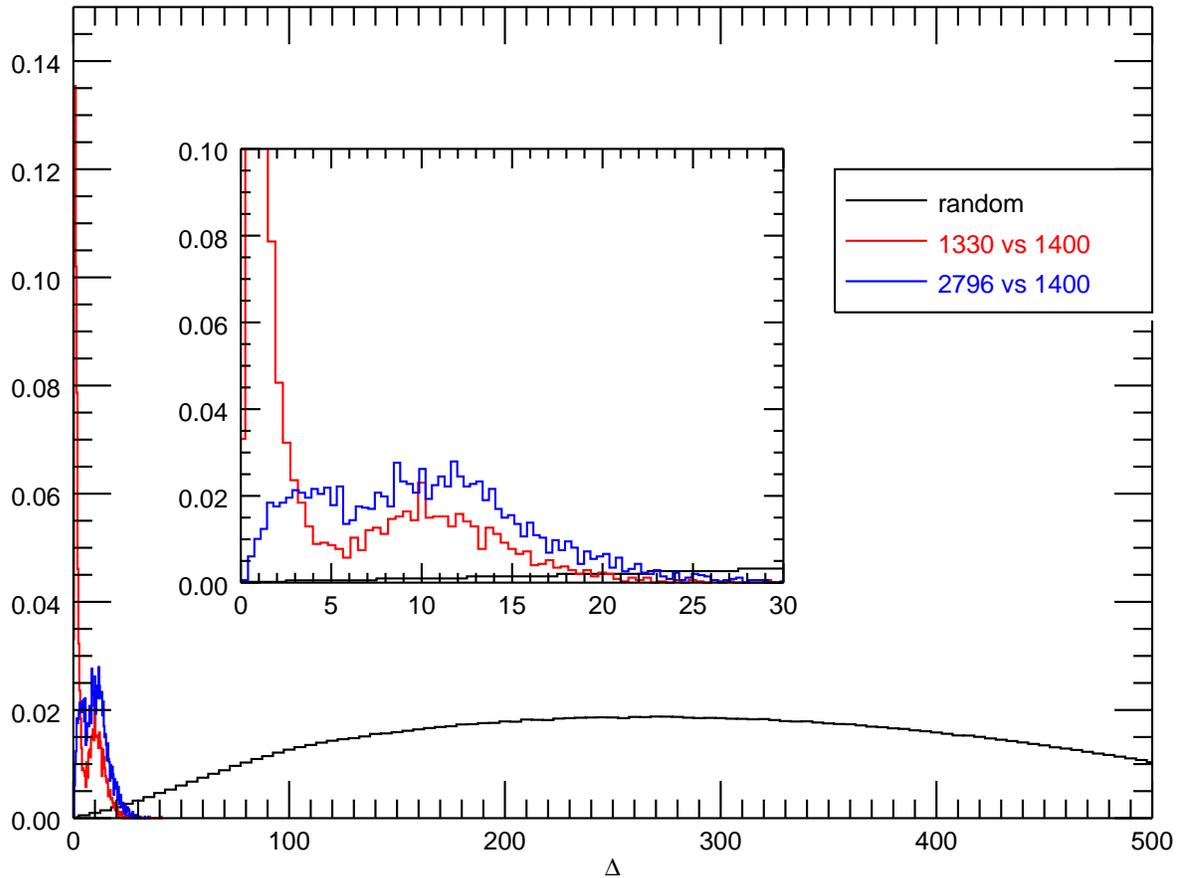}
    \caption{Statistical significance test of $\Delta$ (or the minimum spatio-temporal distance) between 5000 pairs of randomly distributed points (repeated 5000 times). The mean distribution plus 3 standard deviations is plotted in black. Any $\Delta$ values above this baseline are considered a statistically significant match (see text for further details). The distribution for $\Delta$ is also shown between the 1330 and 1400 \AA\ channel detections (red) and the 2796 and 1400 \AA\ channel detections (blue).}
    \label{fig:dst}
\end{figure}

Initially, 3471 small-scale brightening events are detected within the 1400 \AA\ observations. For a given 1400 \AA\ detection, the spatiotemporal distance $\Delta$ is calculated to all detections in the 1330 \AA\ channel. The detection with minimum $\Delta$ is then paired to that 1400 \AA\ detection. This process is repeated over all initial 1400\AA\ detections, and the same procedure pairs 1400\AA\ with 2796 \AA\ detections. We then define an upper limit threshold to $\Delta$ by first defining two groups of 5000 points with random x-, y- and t-positions within a cube of equivalent size to that of the IRIS data set. The $\Delta$ between each corresponding point of these two groups is determined along with its mean and standard deviation. The mean distribution of $\Delta$ plus 3 standard deviations is plotted in black in figure \ref{fig:dst}. The $\Delta$ values between the paired events in 1330 \AA\ vs 1400 \AA\ are then plotted in red while the paired events in 2796 \AA\ vs 1400 \AA\ are plotted in blue. A zoomed-in view is provided in the center of the plot. Any $\Delta$ values that lie above the black `random' baseline are considered significant while those that lie below are discarded. It is clear from the plot that the vast majority of matched detections lie above this baseline with the 1330 \AA\ values crossing at $\Delta\approx20$ and 2796 \AA\ crossing at $\Delta\approx23$. Therefore we choose $\Delta \leq 20$ as a $>3$-sigma confidence threshold. Since we require this threshold to apply across three channels, the true confidence level is actually considerably higher, and we have a high certainty that the events are real since they exist in all channels at a similar position and time.

%The event-by-event distance criterion matches these 1400 \AA\ events with those from the two other wavelengths, reducing the over-compensated number of 1330 \AA\ and 2796 \AA\ detections to equal those of 1400 \AA. 
In tests on synthetic data, \paperi\ defines accurate matches with synthetic events as those with $\Delta \leq 10$ and therefore those with $\Delta>10$ are considered spurious. The larger value of $\Delta \leq 20$ based on the random distribution study of figure \ref{fig:dst}, is reasonable in application to real data considering:
\begin{itemize}
    \item the temporal cadence of this study's data sets are double that of the data set from \paperi 
    \item motions, whether bulk plasma motions or MHD waves, are subject to magnetic field line orientation which may differ between solar layers
    \item possible differential thermal drift, and that these are multi-wavelength and therefore multi-thermal events
\end{itemize}
%A non-arbitrary cut-off - above which events are considered unrelated - is accordingly required. 
%This cut-off is established 

\begin{deluxetable*}{ccccccc}[t]
\tablecaption{The detection threshold values and number of initial detections for all three channels. Note that the thresholds are adjusted so that the number of detections in 1330 \AA\ and 2796 \AA\ match that of 1400 \AA\ plus 10\% \label{tab:study_3_thresh}
}
\tablehead{
\colhead{} & 
\colhead{1400} & 
\colhead{1400} &
\colhead{1330} &
\colhead{1330} &
\colhead{2796} &
\colhead{2796}\\
\colhead{} &
\colhead{threshold} &
\colhead{detections} &
\colhead{threshold} &
\colhead{detections} &
\colhead{threshold} &
\colhead{detections}\\
\colhead{} &
\colhead{($n\sigma$)} &
\colhead{} &
\colhead{($n\sigma$)} &
\colhead{} &
\colhead{($n\sigma$)}
}
\startdata
\threshlo & 9.00 & 5961 & 4.05 & 6547 & 1.86 & 6591\\
\threshhi & 7.00 & 3471 & 5.13 & 3835 & 2.15 & 3890\\
\enddata
\tablecomments{Threshold values are the standard deviation over time ($\sigma$) multiplied by a constant $n$ of the spatially varying threshold.
}
\end{deluxetable*}

\subsection{Power-law Fitting}
\label{sec:power}

The occurrence rate of flares, plotted as a function of total energy, follow power law distributions. These power law distributions indicate an underlying phenomenon of self-organized criticality \citep{Lu_1991} of magnetic fields, whereby a power law index of $\alpha\leq2$ suggests that large-scale, maximum-energy events dominate impulsive heating \citep{Parnell_2008,Krucker_1998}, while a small-scale event dominance would require $\alpha>2$. 
%It is unclear whether this study's observations are in fact nano-flares. Nevertheless, this study's $\alpha$ results suggest the latter. 

The most common method of extracting power-laws, if the data follows such a distribution at all,  is to simply plot a histogram of the data and fit it to a straight line. This basic method is plagued with bias as the gradient of the line fit changes dramatically based on the type of fitting, the behaviour and fluctuations of the tail end of the distribution, and therefore the lower limit for fitting. This is typical of most empirical phenomena. \cite{ryan_2016} demonstrate in their power-law study of X-ray flare flux that flares above a particular ``rollover'' level follow a power law, but those that lie below this level deviate from power-law behaviour. These deviations at the lower end of event distributions - whether the property being fitted is the events' frequency, flux, area or any other - are typically due to under-sampling as a result of spatiotemporal resolution limits, signal-to-noise limits, or background variation \citep{li_2013}. 
There are many approaches to address this undersampling, for example, by using a skew-Laplace distribution rather than a single power law \citep{Wang_2015}. However, this method assumes the density function of the under-recorded (lower-x) side of the data distribution is based on their gradient, which in turn introduces its own host of problems. 
Least-squares regression fitting can provide erroneous estimates for power-law distributions and often fail to discern whether or not the data obey a power law partially due to the assumption of negligible error in independent variables \citep{Cifarelli_1988}. 

Maximum Likelihood Estimation (MLE) methods are among several standard statistical techniques for generating parametric fittings of empirical data \citep{D'Huys_2016}. The Markov chain Monte Carlo (MCMC) method is also a viable alternative for fitting power laws to solar observational data by robustly analysing likelihood functions based on numerous local maxima \citep{Ireland_2015} i.e. `poorly-behaved' and complex data sets. However, for well-behaved likelihood functions, MLE and \cite{Clauset_2007}'s Kolmogorov-Smirnov statistical approach to goodness-of-fit tests provide similar results without the higher computational costs. The maximum of the likelihood function can provide the most likely value of the power-law exponent, $\alpha$, by implementing a $x_{min}$ parameter denoting the lower limit above which the data obey a power law. The question then shifts to how best to estimate this lower limit. \cite{Clauset_2007} present a statistical framework for quantifying power-law behaviour in empirical data by first estimating the lower-bound of the scaling region, $x_{min}$, and scaling parameter, $\alpha$. Kolmogorov-Smirnov statistics \citep{Press_book_1992}, \cite{Clauset_2006}'s complementary cumulative distribution function, scale invariance and binomial regression model are employed to provide objective evidence for whether or not a particular data distribution follows a power law. In this work, we follow \cite{Clauset_2007}, and calculate uncertainties on the power law fitting by recording the variation of the estimated $x_{min}$ and $\alpha$ over 500 random samples of the input data.

%------------------------------------------- 
%Results & Discussion section 
%-------------------------------------------

\section{Results \& Discussion}\label{sec:study_3_results}

\begin{figure}
    \centering
    \includegraphics[trim={1cm 2cm 2cm 2cm},clip,width=\textwidth]{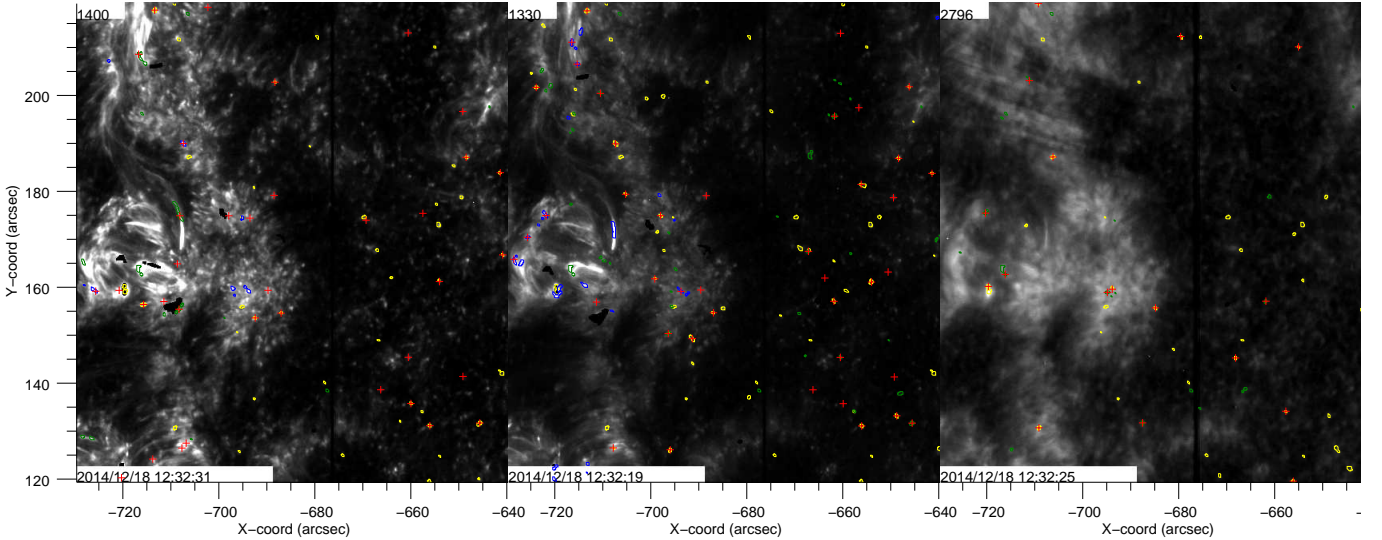}
    \caption{IRIS 1400 \AA\ (left), 1330 \AA\ (center), and 2796 \AA\ (right) slit-jaw images of the ROI taken at approximately 12:32 UT. All detections from the same time frame are over-plotted for each channel. Yellow contours represent brightenings with $N_{frag}\le2$,  green represent $N_{frag}=3$ or $N_{frag}=4$, and blue represents $N_{frag}\ge5$. Red crosses represent events detected across all wavelengths (see text for details). The red crosses without a contour are due to the very small size of the contoured region within this time frame, and some red crosses don't appear in all three images due to a delay of at least one time frame between some coaligned detections. An animation of this figure is available. The video begins on Dec. 18th 2014 at approximately 12:32 UT, and ends on the same day at 13:05 UT.}
    \label{fig:image_comp}
\end{figure}

The number of detections that are coaligned and contemporaneous to all three channels is 2377. Some of these detections are shown in figure \ref{fig:image_comp}, which shows the first frame of the time series in all wavelengths. Yellow contours represent brightenings with $N_{frag}\le2$,  green represent $N_{frag}=3$ or $N_{frag}=4$, and blue represents $N_{frag}\ge5$. Those events detected across all three channels are marked by red crosses. Of these 2377 events within a $90\arcsec\times100\arcsec$ FOV over the course of $\sim34.5$ minutes, 1804, 1759 and 1787 remain un-fragmented or fragment to only two distinct regions at least once during their lifetime (\nfrag$\le2$) for 1400 \AA, 1330 \AA\ and 2796 \AA, respectively. This yields an event density of $9.85\times10^{-5}$ arcsec$^{-2}$ s$^{-1}$, $9.60\times10^{-5}$ arcsec$^{-2}$ s$^{-1}$ and $9.76\times10^{-5}$ arcsec$^{-2}$ s$^{-1}$ for 1400 \AA, 1330 \AA\ and 2976 \AA, respectively, as can be seen in table \ref{tab:study_3_results}. This result contrasts with the result from \paperi's QS observations with an event density of $3.96\times10^{-4}$ arcsec$^{-2}$ s$^{-1}$. This is likely due to the structure within the field of view, and the nature of the thresholding method, whereby active regions (ARs) are, by definition, filled with large, energetic events. These events are typically long-lived and therefore affect the standard deviation of the data set over time, which in turn increases the significance threshold over which events are detected. This results in diminished detections of fainter events within an AR compared to QS regions.

\begin{figure*}
    \centering
        \begin{tikzpicture}
        \node[anchor=south west,inner sep=0] (image) at (0,0) {\includegraphics[trim={0cm 0cm 0cm 0cm},clip,width=\textwidth]{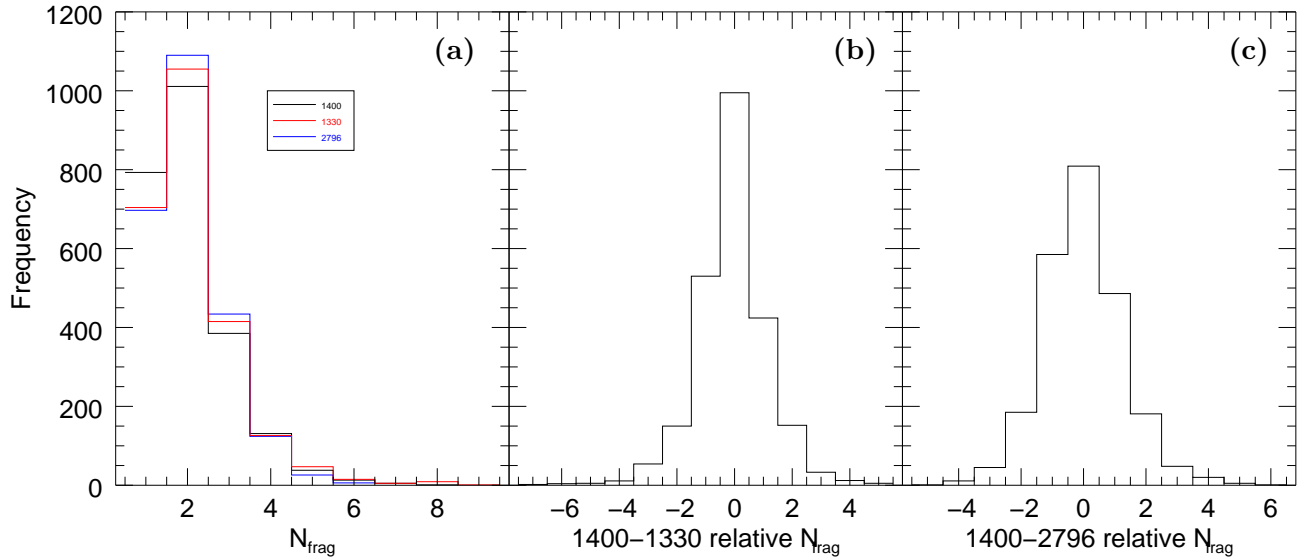}};
        \begin{scope}[
        x={(image.south east)},
        y={(image.north west)}
    ]
        \node [black, font=\bfseries] at (0.34,0.90) {(a)};
        \node [black, font=\bfseries] at (0.635,0.90) {(b)};
        \node [black, font=\bfseries] at (0.925,0.90) {(c)};
    \end{scope}
    \end{tikzpicture}  
    
    \caption{Histograms of fragmentations during the lifetime of detections. (a) directly compares the distribution of \nfrag\ values for 1400 \AA\ (black), 1330 \AA\ (red), and 2796 \AA\ (blue) detections. $N_{frag}$ values range between 1 - 8 for both 1440 \AA\ detections, 1-9 for 1330 \AA\ detections, and 1 - 6 for 2796 \AA\ detections. (b) and (c) show the relative \nfrag\ value difference between each coaligned and contemporaneous event. See text for further details.}
    \label{fig:study_3_nfrag_histogram}
\end{figure*}

The remainder of the 1400 \AA\ events consist of $3\le N_{frag}\le8$, 1330 \AA\ events consist of $3\le N_{frag}\le9$, and the 2796 \AA\ events consist of $3\le N_{frag}\le6$, as can be seen in figure \ref{fig:study_3_nfrag_histogram}(a). This result also contrasts with those of \paperi, whereby $N_{frag}$ reaches a value of 12. This suggests that the events detected in this study are typically less complex and fragmented as those of the previous study. Figures \ref{fig:study_3_nfrag_histogram}(b) and \ref{fig:study_3_nfrag_histogram}(c) show histograms of the relative difference between $N_{frag}$ values of each matched event for 1400 vs 1330 and 1400 vs 2796, respectively. Approximately half of all matched events fragment an equal amount across each wavelength (i.e. a relative $N_{frag}$ value of 0). A mostly symmetric distribution is apparent for both of these histograms with slight weighting towards the negative end i.e. 1330 and 2796 \AA\ detections tend to be less fragmented than their 1400 \AA\ counterpart.

\begin{figure*}
    \centering
        \begin{tikzpicture}
        \node[anchor=south west,inner sep=0] (image) at (0,0) {\includegraphics[trim={0cm 0.5cm 0.1cm 0.7cm},clip,width=1\textwidth]{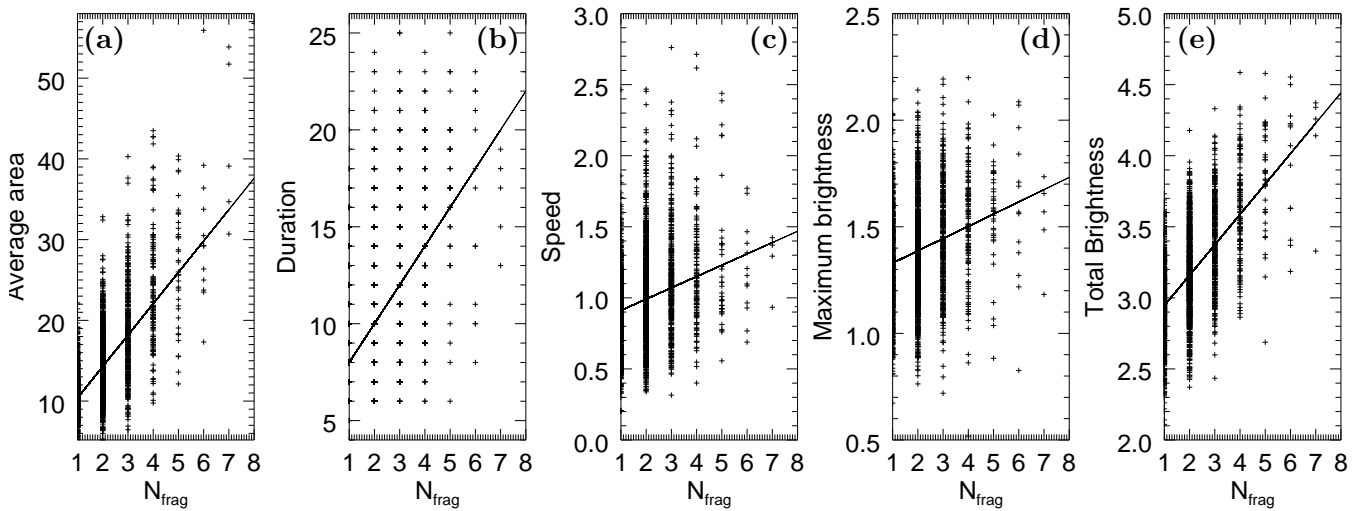}};
        \begin{scope}[
        x={(image.south east)},
        y={(image.north west)}
    ]
        \node [black, font=\bfseries] at (0.07,0.90) {(a)};
        \node [black, font=\bfseries] at (0.36,0.90) {(b)};
        \node [black, font=\bfseries] at (0.56,0.90) {(c)};
        \node [black, font=\bfseries] at (0.76,0.90) {(d)};
        \node [black, font=\bfseries] at (0.875,0.90) {(e)};
    \end{scope}
    \end{tikzpicture}    
    
    \caption{Scatter plots showing relationships between the $N_{frag}$ parameter and (a) average area (pixels), (b) duration (frames), (c) average speed (pixel shift per frame), (d) maximum brightness ($log_{10}$[DN s$^{-1}$]), and (e) total brightness ($log_{10}$[DN s$^{-1}$]). Diagonal black lines represent a least-absolute-deviation fitting procedure (see text).}
    \label{fig:study_3_nfrag_comp_with_other_stats}
\end{figure*}

Figure \ref{fig:study_3_nfrag_comp_with_other_stats} shows plots of (a) average area, (b) duration, (c) average speed, (d) maximum brightness and (e) total brightness as a function of the $N_{frag}$ parameter for the 1400 \AA\ detections. We fit these scattered points to a straight line using a least-absolute-deviation (LAD) fitting procedure, (the solid lines) presenting gradients of 3.89, 2.00, 0.08, 0.06, and 0.21 respectively. These gradients imply a correlation between the $N_{frag}$ parameter and average area as well as duration, which suggest that larger and longer-lasting events are more likely to fragment during their lifetime. No apparent relationship exists between $N_{frag}$ and average speed, maximum brightness or total brightness. These trends also generally apply to both 1330 \AA\ and 2796 \AA\ detections.

\begin{figure*}
    \centering
        \begin{tikzpicture}
        \node[anchor=south west,inner sep=0] (image) at (0,0) {\includegraphics[trim={0cm 0cm 0cm 0cm},clip,width=\textwidth]{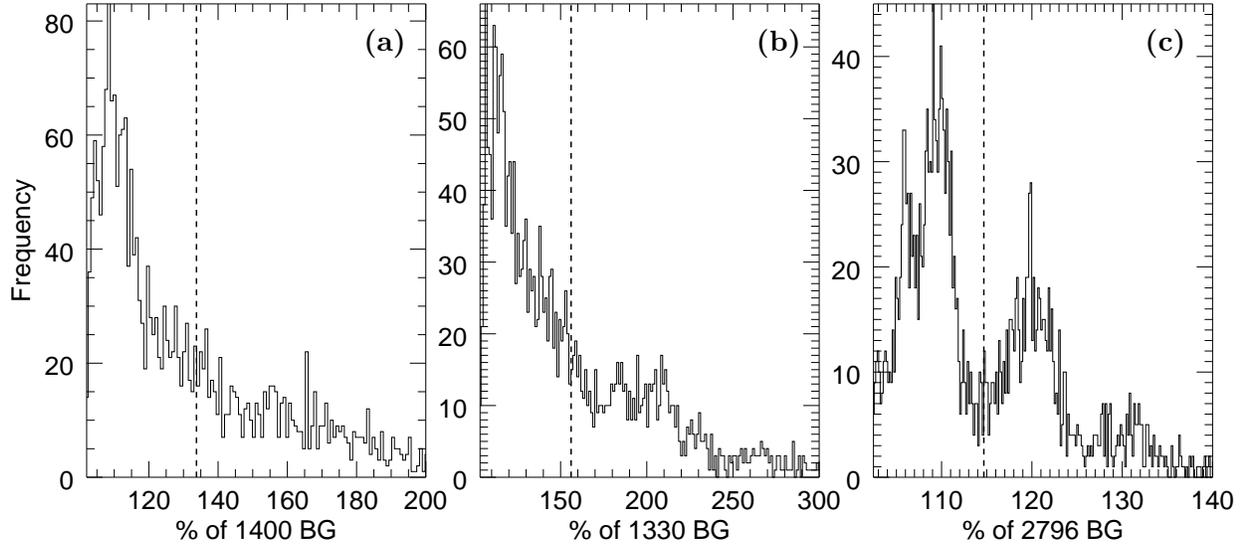}};
        \begin{scope}[
        x={(image.south east)},
        y={(image.north west)}
    ]
        \node [black, font=\bfseries] at (0.31,0.90) {(a)};
        \node [black, font=\bfseries] at (0.60,0.90) {(b)};
        \node [black, font=\bfseries] at (0.885,0.90) {(c)};
    \end{scope}
    \end{tikzpicture}  
    
    \caption{Histograms of the percentage values of events' maximum brightness relative to their average background intensity (in DNs$^{-1}$). (a) displays the relative intensity values of detections in 1400 \AA, (b) displays those of 1330 \AA, and (c) displays those of 2796 \AA. Vertical dashes lines represent average values of 134\%, 156\% and 115\%, respectively. ``BG" denotes background intensity.}
    \label{fig:study_3_maxb_vs_bg}
\end{figure*}

As a reference, figure \ref{fig:study_3_maxb_vs_bg} displays the percentage values of detected maximum brightness (in DNs$^{-1}$) from each wavelength relative to their average background intensity (also in DNs$^{-1}$). (a) displays the relative intensity values of detections in 1400 \AA, (b) displays those of 1330 \AA, and (c) displays those of 2796 \AA. The vertical dashes lines represent the average relative intensity values of 134\%, 156\% and 115\%, respectively.

We restrict the remainder of the analysis to those detections with $N_{frag}\le2$. This $N_{frag}$ parameter is an interesting property which demands further attention in a future study. 

\begin{deluxetable*}{cccc}
\tablecaption{Comparison of the number of detections, event densities and maximum $N_{frag}$ values for all three channels. \label{tab:study_3_results}
}
\tablehead{
\colhead{Channel} & 
\colhead{Detections} &
\colhead{Event density} &
\colhead{Max. $N_{frag}$}\\
\colhead{\AA} &
\colhead{$N_{frag}\le2$} &
\colhead{arcsec$^{-2}$ s$^{-1}$} &
\colhead{}\\
}
\startdata
1400 & 1804 & $9.85\times10^{-5}$ & 8\\
1330 & 2070 & $9.60\times10^{-5}$ & 9\\
2796 & 2020 & $9.78\times10^{-5}$ & 6\\
\enddata
\tablecomments{Event densities are based on the number of detections with $N_{frag}\le2$ (see text).
}
\end{deluxetable*}

\subsection{Property analysis}

\begin{figure*}
    \centering
    \includegraphics[trim={0cm 0cm 0cm 0cm},clip,width=\textwidth]{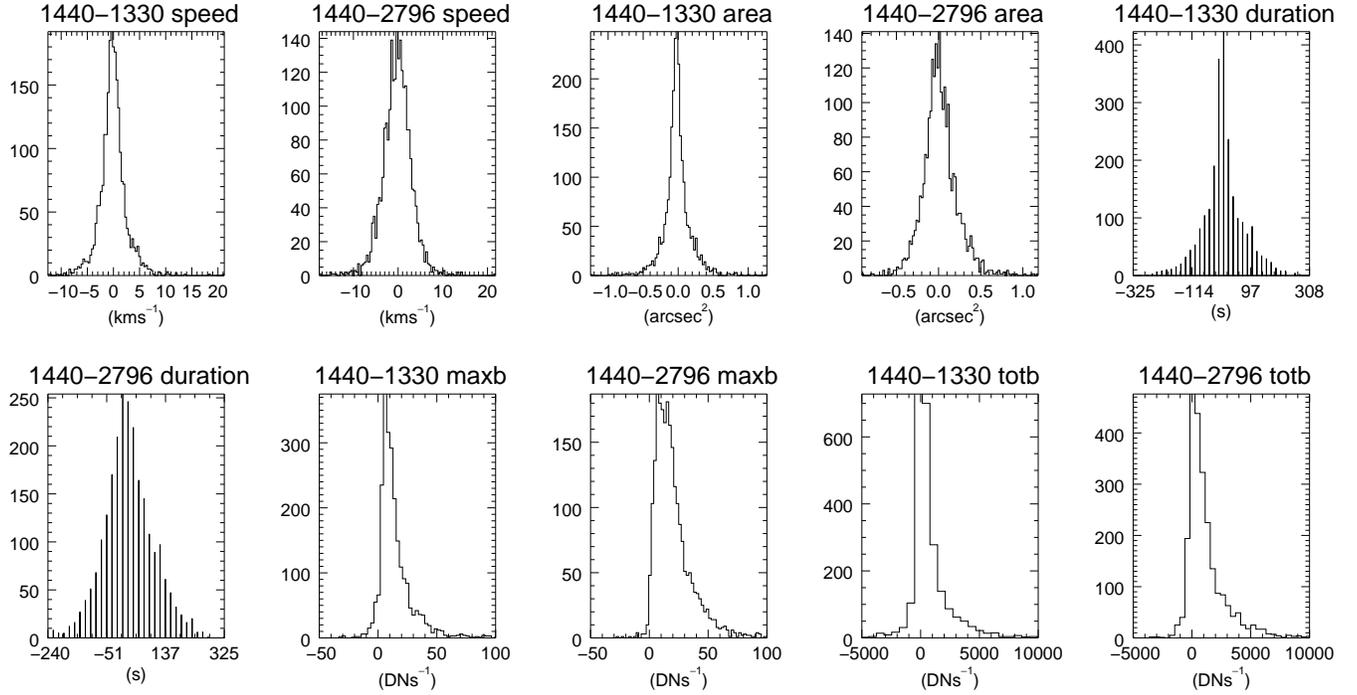}
    \caption{Difference histograms comparing the characteristics of each paired 1400/1330 and 1400/2796 event, including average speed, average area, duration, maximum brightness and total brightness (see plot titles).}
    \label{fig:study_3_histograms}
\end{figure*}

Figure \ref{fig:study_3_histograms} shows histograms of the difference between various characteristics of the paired 1400/1330 and 2796/1330 detections: namely speed (average over time for each event), area, duration, maximum brightness and total brightness (as indicated in the plot titles). A mostly symmetrical distribution centered at zero is present when comparing speed, area, and duration, regardless of wavelength. 
It is evident that the maximum and total brightness differences peak above zero with a slight weight towards the positive end, suggesting that a disproportionate number of 1400 \AA\ detections are brighter than their 1330 and 2796 \AA\ counterparts.

Additionally, the mean duration, maximum brightness and total brightness are recorded in table \ref{tab:study_3_stats}. The mean of the average area and average speed are also recorded i.e. the average area and average speed of each event over its lifetime is recorded, and table \ref{tab:study_3_stats} displays the mean of these average area and average speed results. 1400 \AA\ and 1330 \AA\ events appear similar, displaying very little difference between their area, speed and duration. 2796 \AA\ detections seem to deviate from these common results, demonstrating lower average area, higher average speeds and shorter durations. 
Large differences are evident when comparing the maximum and total brightnesses of each wavelength channel. However, as explained in section \ref{sec:study_3_detection_method}, intrinsic differences in the effective area of each channel and therefore the number of photons captured by the CCDs differ, whereby the 1400 \AA\ channel possesses the greatest effective area. Table \ref{tab:study_3_stats} compares the mean of each characteristic of all events from each wavelength channel. The mean total brightness result of the 1400 \AA\ detections appear lower than that of the events in \paperi\ by a factor of $\sim3$. The mean of the average speeds from each pass-band (7.22-7.36\kms) also appear slower than that of \paperi's observations (9.03 km s$^{-1}$). However, the mean average area of the 1400 \AA\ detections (0.412 arcsec$^2$) is larger than \paperi's 0.32 arcsec$^2$.

These results share several similarities with those of \cite{nhalil_2020}. Although \cite{nhalil_2020} focus on spectral observations of coronal upflows, they also employ a method of tracking dynamic bright spots in IRIS 1400 \AA. These bright dots have an average size of 0.3 Mm$^{2}$ and durations mostly $<200$s, compared to this study's average area of 0.22 Mm$^{2}$ ($\sim0.41$ arcsec$^{2}$) and where $\sim70\%$ of events have a lifetime $\leq200$s. However, over 90\% of their dots have a lifetime $\leq100$s, whereas this study focuses on detections with lifetimes of at least 85s and nearly all detections last longer than 100s.
Similarly, \cite{Tiwari_2019}'s rigorous Hi-C-2.1 \citep{Rachmeler_2019} study of ``dot-like" brightening events at the base of an AR filament system are of interest. While these dot-like events are far larger ($\sim2\arcsec$) than those detected in this study or that of \paperi, their lifetimes lie well within the capabilities of this detection code, whereby \paperi\ detects small-scale events as short-lived as 40s. \cite{Tiwari_2019} also observe elongated brightening events within small-scale magnetic loops (with lengths of $\sim2.5\arcsec\approx$1.8 Mm and lifetimes of 25-230s) as well as small-scale jet-like events (with lengths of $\sim5\arcsec\approx3.6$ Mm and lifetimes of 50-300s), the latter of which are similar to ``jetlet" observations by \cite{Raouafi_2014} and \cite{Panesar_2019}. 

\cite{Panesar_2019} adds that the typical width of these jetlets are 300-850 km, demonstrating clear disparities between the lengths and widths of such structures. Our detection code records information on the elongation and orientation of detected structures as a function of time, although this information is not used in this study. In principle, such information can be valuable to categorize these results into different types of events, but would involve further work that is beyond the scope of this study. Therefore, within the limits that we impose on our method, the features detected and analysed by \cite{Raouafi_2014}, \cite{Panesar_2019}, \cite{nhalil_2020}, and \cite{Tiwari_2019} can all be readily detected by our method.

% Dividing these by their Effective Area (EA) (in cm^2) = 2231.44, 4024.68 and 1.24x10^5!! 2796 is extremely bright if we're to take this as is!!)!!

\subsection{Power-law analysis}

\begin{deluxetable*}{ccccccc}[t]
\tablecaption{Comparison of the mean characteristics of all detected events from each wavelength channel following the $\Delta=20$ criterion. Bracketed values are standard deviations. \label{tab:study_3_stats}
}
\tablehead{
\colhead{Channel} & 
\colhead{Av. area} &
\colhead{Av. Speed} &
\colhead{Duration} &
\colhead{Max. Brightness} &
\colhead{Tot. Brightness}\\
\colhead{(\AA)} &
\colhead{acsec$^{2}$} &
\colhead{(km s$^{-1}$)} &
\colhead{(s)} &
\colhead{(DN s$^{-1}$)} &
\colhead{(DN s$^{-1}$)}
}
\startdata
1400 & 0.412 (0.169) & 7.29 (2.51) & 180.02 (63.77) & 28.81 (18.21) & 2237.52 (2909.45) \\
1330 & 0.414 (0.168) & 7.22 (2.43) & 180.65 (63.14) & 13.01 (8.16) & 1022.87 (1587.56)\\
2796 & 0.386 (0.144) & 7.36 (2.52) & 160.66 (52.87) & 7.50 (4.43) & 594.39 (630.94)\\
\enddata
\tablecomments{Intensities are measured in DN (IRIS counts or Data Number) s$^{-1}$.
}
\end{deluxetable*}
%2.24x10^5 1400 mean area
%2.27x10^5 1330 mean area
%2.11x10^5 2796 mean area

\begin{figure*}
    \centering
    \begin{tikzpicture}
        \node[anchor=south west,inner sep=0] (image) at (0,0) {\includegraphics[trim={0cm 0cm 0cm 0cm},clip,width=0.95\textwidth]{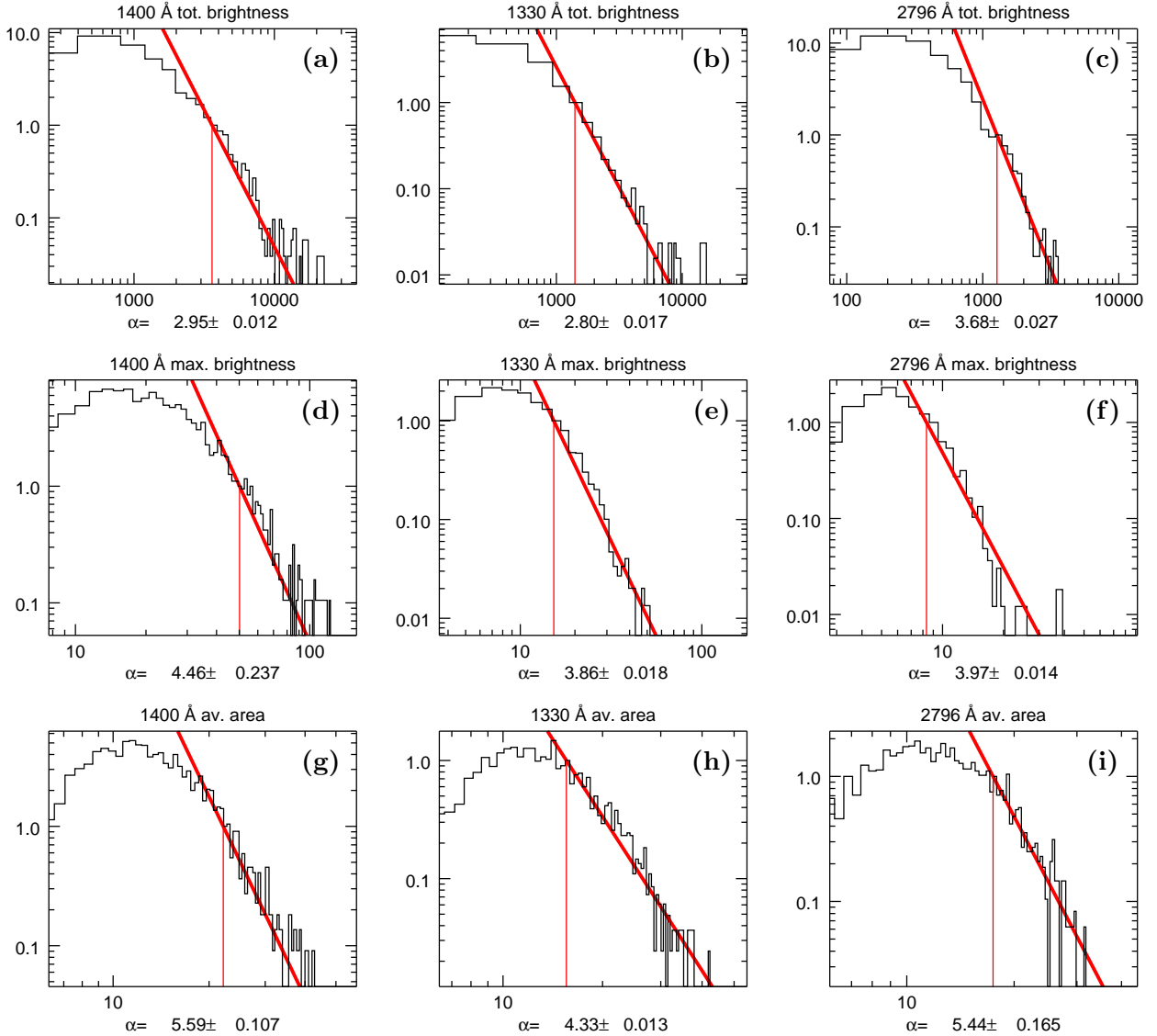}};
        \begin{scope}[
        x={(image.south east)},
        y={(image.north west)}
    ]
        \node [black, font=\bfseries] at (0.29,0.93) {(a)};
        \node [black, font=\bfseries] at (0.62,0.93) {(b)};
        \node [black, font=\bfseries] at (0.95,0.93) {(c)};
        \node [black, font=\bfseries] at (0.29,0.60) {(d)};
        \node [black, font=\bfseries] at (0.62,0.60) {(e)};
        \node [black, font=\bfseries] at (0.95,0.60) {(f)};
        \node [black, font=\bfseries] at (0.29,0.27) {(g)};
        \node [black, font=\bfseries] at (0.62,0.27) {(h)};
        \node [black, font=\bfseries] at (0.95,0.27) {(i)};
    \end{scope}
    \end{tikzpicture}

    \caption{Distributions and power-law indices for event total brightness in (a) 1400 \AA, (b) 1330 \AA\ and (c) 2796 \AA, maximum brightness in (d) 1400 \AA, (e) 1330 \AA\ and (f) 2796 \AA, and average area in (g) 1400 \AA, (h) 1330 \AA\ and (i) 2796 \AA. The diagonal red line indicates the derived power-law gradient using \cite{Clauset_2007}'s method. Vertical red line indicates the lower bound of the scaling region, above which the distribution obeys the power law.}
    \label{fig:power_law_study_3}
\end{figure*}

The total brightness of the events detected in each channel are fitted to power-laws distributions using the method described in \cite{Clauset_2007}. Figure \ref{fig:power_law_study_3} (a), (b) and (c) display the results of applying this method to the 1400 \AA\, 1330 \AA\ and 2796 \AA\ events' total brightness, respectively. We find that these distributions display scaling parameters of $\alpha=2.95\pm0.012$, $\alpha=2.80\pm0.017$ and $\alpha=3.68\pm0.027$, respectively. Figure \ref{fig:power_law_study_3} (d), (e) and (f) demonstrate that the maximum brightness distributions also follow a power-law with index values of $4.46\pm0.023$, $3.86\pm0.018$ and $3.97\pm0.014$, respectively, for 1400 \AA, 1330 \AA\ and 2796 \AA, respectively. The same is also true for average area distributions with figure \ref{fig:power_law_study_3} (g), (h) and (i) showing power-law indices of $5.59\pm0.107$, $4.33\pm0.013$ and $5.44\pm0.165$ for 1400 \AA, 1330 \AA\ and 2796 \AA, respectively.

These total brightness $\alpha$ values are far higher than those determined in other observational studies of transient events (e.g. \cite{shimizu_1995}, \cite{Berghmans_1998}, \cite{Berghmans_2001}, \cite{Hannah_2008}), with a varied $1.5\ge\alpha\ge2.2$ range. It is possible that these studies under-count the number of events within their ROIs, particularly those at lower energies. While the large AR within this study's FOV may affect the spatially-varying threshold above which events are detected, the filtering method successfully detects very small (with a mean area of $\sim0.4$ arcsec$^2$), dim events. Additionally, these studies are conducted over a wide range of cadences, passbands and instrument resolutions. This inconsistency may also result in the detection of different phenomena, driven by different mechanisms and therefore may not follow the same power law distributions.

We draw comparisons in \paperi\ between the properties of the detected events and those of other small-scale phenomena, such as Ellerman Bombs \citep{ellerman}. While these events typically manifest in H$\alpha$ lines, their size and duration are not dissimilar to this study's observations. \cite{Georgoulis_2002} demonstrate that Ellerman Bomb distributions exhibit power-law shapes not only for energy release (power law index of $\sim2.1$) but also for duration ($\alpha\simeq2.22$) and area ($\alpha\simeq2.44$) using least-squares fitting.
The average area power-law $\alpha$ values of this study are far higher than those of \cite{Georgoulis_2002} by at least a factor of 2. Additionally, Ellerman Bomb temperatures do not exceed 10,000 K, while this study's events demonstrate Si {\sc{iv}} signatures which lie within $5000$-$1.5\times10^{5}$K. This temperatures range overlaps with that of IRIS Bombs \citep{Peter_2014}, but our detections can exhibit far larger areas with widths of $\sim5\arcsec$ i.e. 30 IRIS pixels compared to (assuming an event circular shape) this study's maximum width of $\sim7\arcsec$ or $\sim4.3$ IRIS pixels.

\cite{nhalil_2020} demonstrate $1.5\leq\alpha\leq2.25$ values for 1-, 2- and 3-IRIS-pixel brightening events. Each event of this study has a volume of at least 5 pixels with a larger range of volume values. Cadences are also set at 60s and 110s with which they conclude that the choice of cadence affects the scaling parameter values. Perhaps the relatively `large' temporal resolution (i.e. a short cadence of $\sim17$s) of this study may account for larger scaling parameter values, whereby total energy output may be more accurately resolved over shorter gaps in time between frames. 

A possible avenue of inquiry would be to cross-analyse a H$\alpha$ ROI where Ellerman Bombs are present with a co-aligned IRIS set of sit-and-stare, multi-wavelength data set. Such an analysis may present reasonable evidence that the small-scale events observed in this study are in fact the TR counterparts to Ellerman Bombs. \cite{Qiu_2000}'s analysis of bright points in 1600 \AA\ reinforces this proposal, suggesting that Ellerman Bombs do exhibit at least some UV emission. 

\begin{figure*}
    \centering
    \begin{tikzpicture}
        \node[anchor=south west,inner sep=0] (image) at (0,0) {\includegraphics[trim={0cm 0cm 0cm 0cm},clip,width=0.95\textwidth]{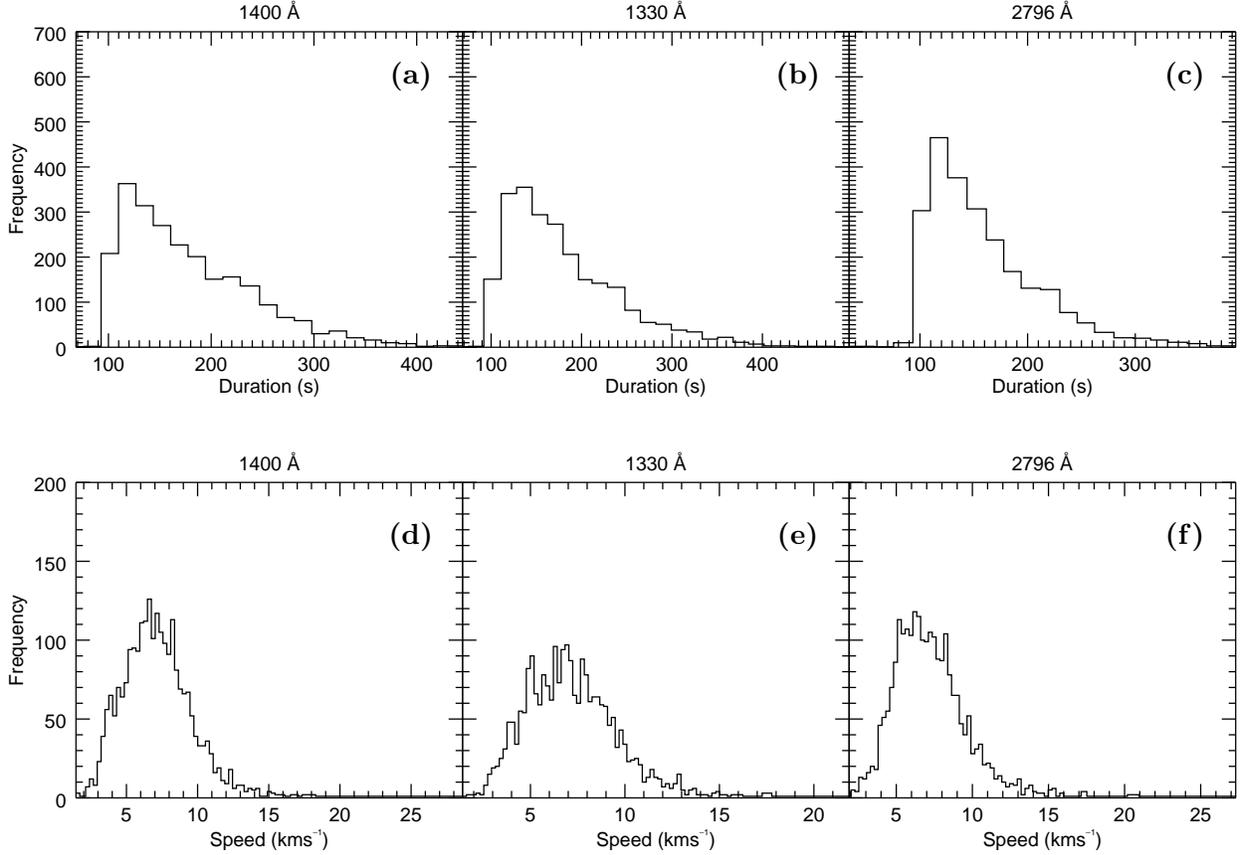}};
        \begin{scope}[
        x={(image.south east)},
        y={(image.north west)}
    ]
        \node [black, font=\bfseries] at (0.34,0.90) {(a)};
        \node [black, font=\bfseries] at (0.64,0.90) {(b)};
        \node [black, font=\bfseries] at (0.94,0.90) {(c)};
        \node [black, font=\bfseries] at (0.34,0.39) {(d)};
        \node [black, font=\bfseries] at (0.64,0.39) {(e)};
        \node [black, font=\bfseries] at (0.94,0.39) {(f)};
    \end{scope}
    \end{tikzpicture}
    \caption{Histograms of event duration for (a) 1400 \AA, (b) 1330 \AA\ and (c) 2796 \AA, and of event speed for (d) 1400 \AA, (e) 1330 \AA\ and (f) 2796 \AA.}
    \label{fig:speed_duration_histograms_study_3}
\end{figure*}

The distribution of these events' duration and speed do not appear to follow a power law. These distributions are plotted in figure \ref{fig:speed_duration_histograms_study_3}. The top row (a-c) show the duration for the three channels, and there is a clear cut-off of detections below $\sim90$s (or 5 frames at $\sim17$s cadence, i.e. our fixed lower limit for detection). The bottom row (d-f) shows the speed distributions. These distributions suggest that the speed of all events are non-zero with minimum speeds of $\sim1.45$\kms, $\sim1.40$\kms\ and $\sim1.89$\kms\ for 1400 \AA, 1330 \AA\ and 2796 \AA, respectively, with all wavelengths sharing a median average speed of $\sim7$\kms. \cite{Huang_2021}'s bright dot effective speeds are of comparable ranges to this study's detections whereby 90\% of their results lie below 15\kms\ (compared to $\sim98\%$ of this study's results). However, $\sim65\%$ of their effective speed values are $\leq1$\kms, while none of this study's speed results lie below 1\kms. It is also unclear from their study whether any of their bright dots remain static throughout their lifetime or not.

If these detections are observations of energetic events (i.e. an initial release of energy via magnetic reconnection or otherwise) then the expected speed distribution would peak at approximately zero with a gradual drop off with faster speeds - it is not necessary to expect heated plasma to move following a reconnection event. However, this is not the case as all detections appear to move. It is possible that these observations are a jet-like signature following an initial reconnection event, or perhaps they are a manifestation of a different type of energetic event. Magnetic reconnection in solar flares are often accompanied by an inflow/outflow of plasma, such as those observed by \cite{Yokoyama_2001} with apparent inflow speeds of 1.0-4.7\kms. Other flare observations demonstrate outflow speeds in the $\sim10^{3}$ \kms\ range \citep{Savage_2012, Wang_2007} although this large discrepancy may be due to the difference in scale (both size and energy output) between large flares and the small-scale detections of this study. 

Finally, it is worth noting that this study's detection method allows for the tracking of an event's centroid over the course of its lifetime and these Plane-of-Sky (PoS) motions (i.e. the proper motions of brightening events) often change direction from one frame to the next. Therefore an analysis of the detections' acceleration may be of interest in a future study, including a statistical analysis of their paths of PoS motion.

\subsection{Motion tracking}
\begin{figure}
    \centering
    \includegraphics[width=0.98\textwidth]{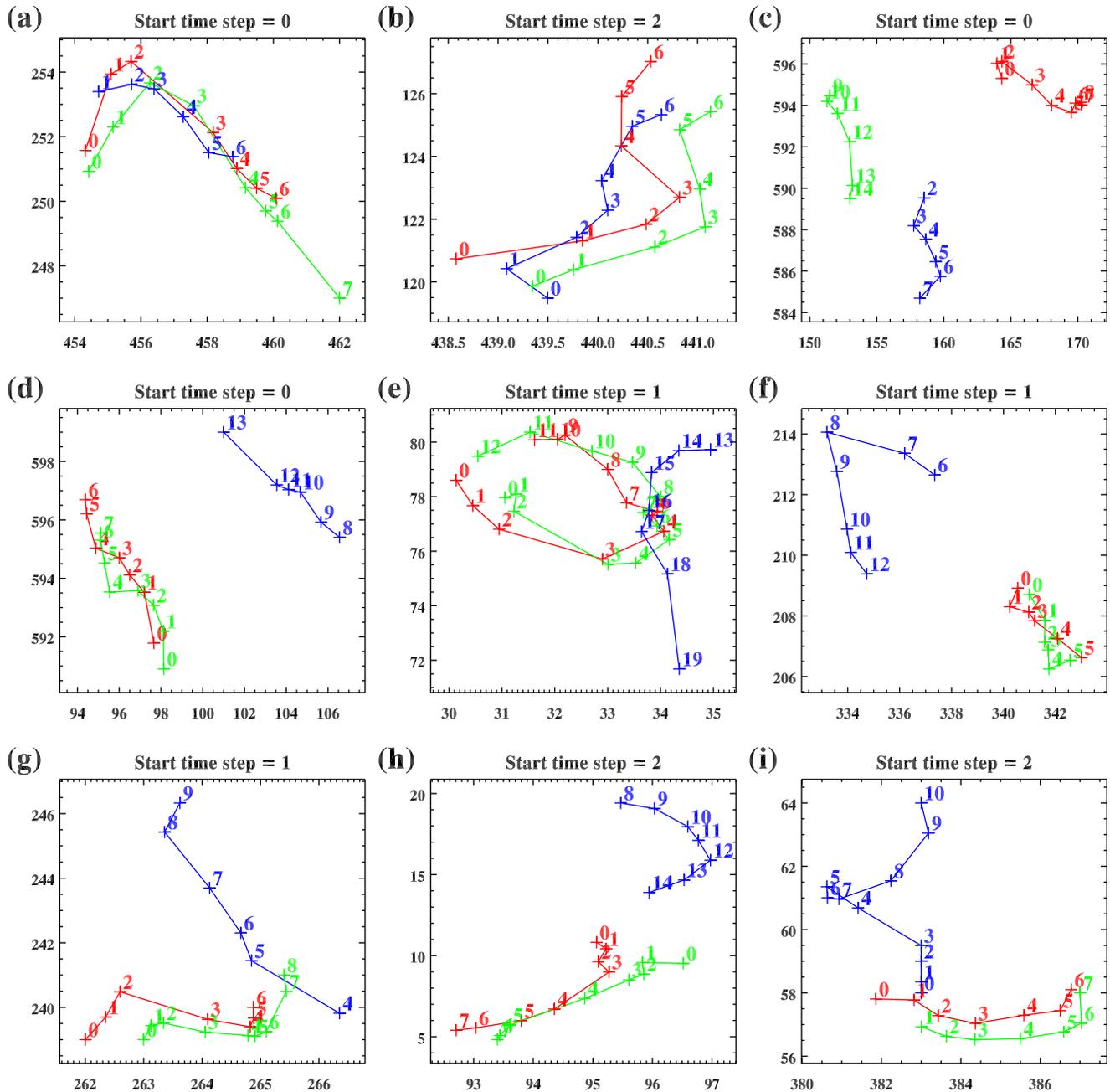}
    \caption{The paths, over time, of 9 example detections for channels 1400 \AA\ (red), 1330 \AA\ (green), and 2796 \AA\ (blue). The numbers next to the plotting symbols refer to the time step for that detection relative to the start time step indicated in the plot titles. The $x$ and $y$ axis are in units of pixels within the datacube.}
    \label{motion}
\end{figure}

One powerful result of our method is the ability to track the centroid of an event over time. This is the weighted mean $x$ and $y$ co-ordinates, weighted by the background-subtracted intensities of pixels contained within the event at each time step. Figure \ref{motion} shows nine selected events that start within the first three time steps of the data set. In each case, we show the time evolution of the centroids for the three channels in different colors, and label each centroid with the time step number. Many of these PoS motion tracks show excellent agreement between all three channels, as shown in figure \ref{motion}a and b. For these cases, there is very high confidence that the same event is being viewed in all channels, and the event is obviously multithermal since the event is seen at the same time in all channels. Some cases show three separated tracks for the three channels, as shown in figure \ref{motion}c. In this case, we cannot be so confident in interpreting the connection between channels. The spatio-temporal separation of the detections between channels gives confidence that the events are connected - the chances of 3 detections occuring by random within such a small volume are very small (see figure \ref{fig:dst}), and there are many events that show this small separation in paths and PoS motion. We therefore conclude that the detections are plasma enhancements (density and/or thermal) that arise from a single underlying event.

Figures \ref{motion}d to i show examples of the most common pattern. In these cases, the 1330 and 1400 \AA\ detections show very similar PoS motions, whilst the 2796 \AA\ detection is separated and/or shows a different trajectory. These events occur frequently, therefore we have confidence that the pattern must be caused by a common underlying event - at least for a proportion of these cases. Figures \ref{motion}d, f, g, and h show the 2796 \AA\ path separated spatially and temporally from the hotter channels by a few pixels/time steps, with a tendency to start a few time steps later. Figure \ref{motion}i shows an example where the detections in all 3 channels start at a similar position and time, but the 2796 \AA\ detection diverges on a different trajectory to the other two channels. This is evidence of different behaviour at different temperatures originating from a single event, and could be interpreted as a cool and hotter jet propagating along different field lines, or in different directions along the same field lines. Figure \ref{motion}e shows a case where the hotter channels describe almost a closed loop, with a subsequent PoS motion in the cooler channel passing close through the mid-point of the loop.

\begin{figure}
    \centering
    \includegraphics[width=0.98\textwidth]{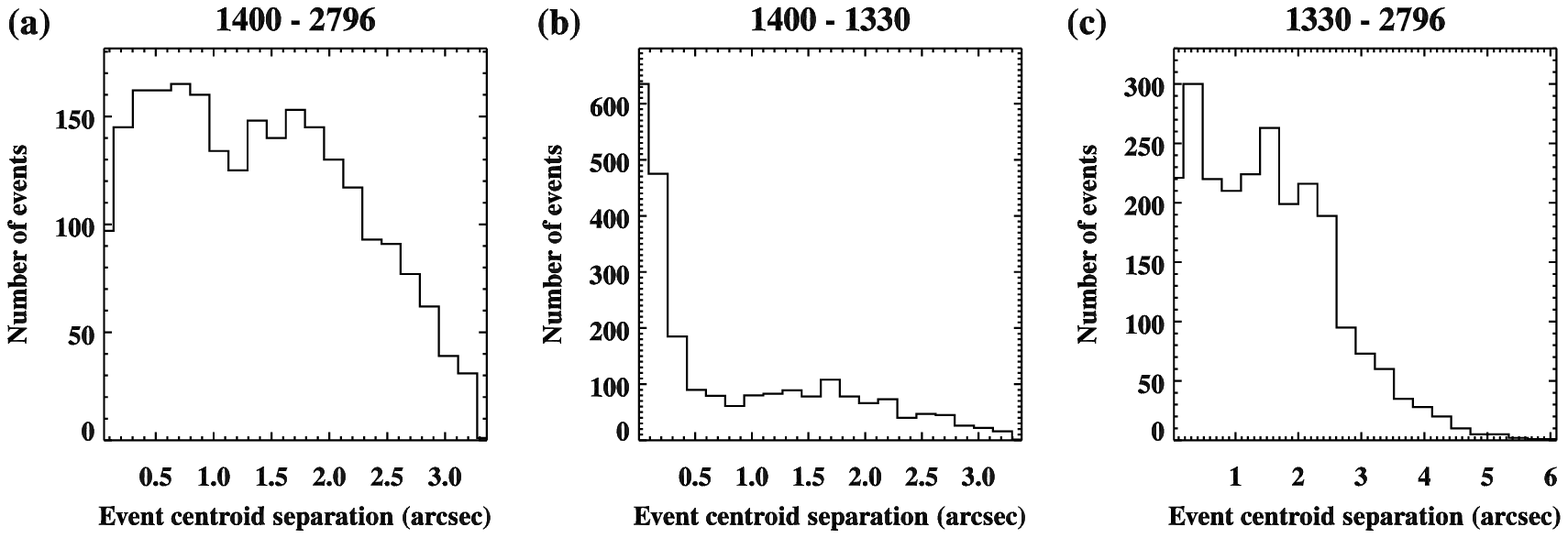}
    \caption{Distribution of the mean spatial distance between events between channels (a) 1400 \AA\ and 2796 \AA, (b) 1400 \AA\ and 1330 \AA, and (c) 1330 \AA\ and 2796 \AA.}
    \label{eventsep}
\end{figure}

Figure \ref{eventsep} shows, for all 2377 detected events, the spatial distance between event centroids between pairs of channels. These distribution of figure \ref{eventsep}b confirms that a large number of events are co-spatial in the 1400 and 1330 \AA\ channel. Figures \ref{eventsep}a and c show that the 2796 \AA\ channel events tend to be at greater distances from events in other channels, with the greatest separation between the 1330 and 2796 \AA\ channels. These spatial distances between a matched event in each wavelength are based on the events' centroids, and we should not necessarily expect these centroids to agree exactly. Whilst centroids may be different, an event observed in one channel may overlap in area with its counterpart in a different channel. It is also worth noting that uncertainty and noise can influence the position of an event's centroid at these scales.

\begin{figure}
    \centering
    \includegraphics[width=0.98\textwidth]{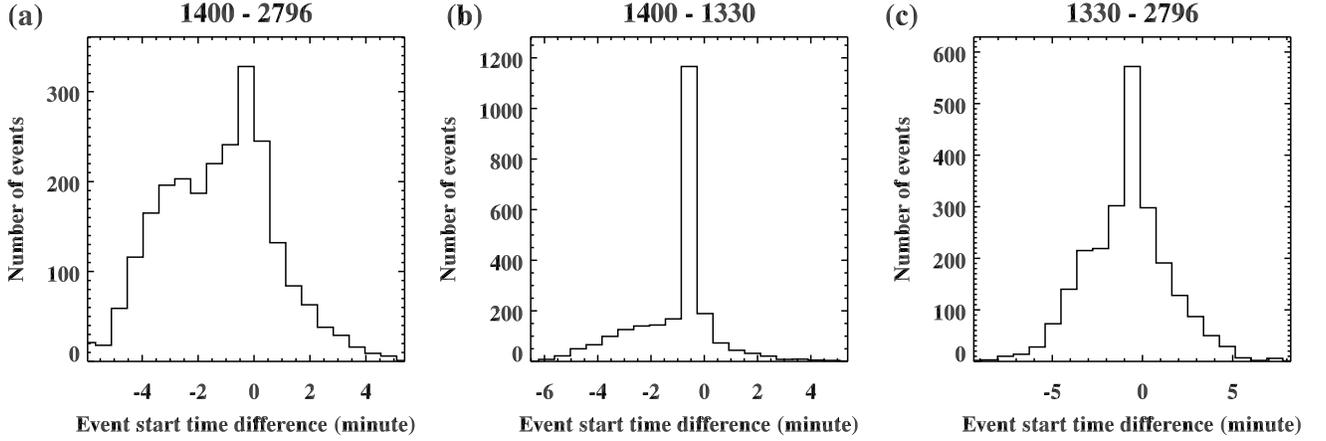}
    \caption{Distribution of the difference in event start times for channels (a) 1400 \AA\ start times minus 2796 \AA\ start times, (b) 1400 \AA\ minus 1330 \AA, and (c) 1330 \AA\ minus 2796 \AA.}
    \label{eventst}
\end{figure}

Figure \ref{eventst} shows the difference in event start times between channels. Figure \ref{eventst}a shows that there is a strong statistical tendency for 1400 \AA\ events to precede the 2796 \AA\ events. There is a weak tendency for 1330 \AA\ events to precede 1400 \AA\ events, although the distribution is strongly peaked near zero. The 1330-2796 \AA\ distribution, shown in figure \ref{eventst}c, is peaked near zero, and is more symmetrical. Thus statistically, events tend to appear in 1400 \AA\ before the other two channels. We note that distributions showing the event median times, and end times, show a similar pattern, thus this pattern is not a result of the sensitivity of the detection method in different channels. This result is therefore important - it suggests that in many events, the event appears first in the hottest channels, before appearing a few time steps later in the cooler channel. The most obvious physical interpretation of this is that the events involve impulsive heating followed by rapid cooling. Another possible interpretation is that the reconnection occurs at a higher elevation (TR), with a subsequent propagating disturbance to the lower atmosphere. The fact that events in the coolest channel (2796 \AA) tend to appear displaced spatially from the hotter channels may support the latter scenario.

\begin{figure}
    \centering
    \includegraphics[width=0.98\textwidth]{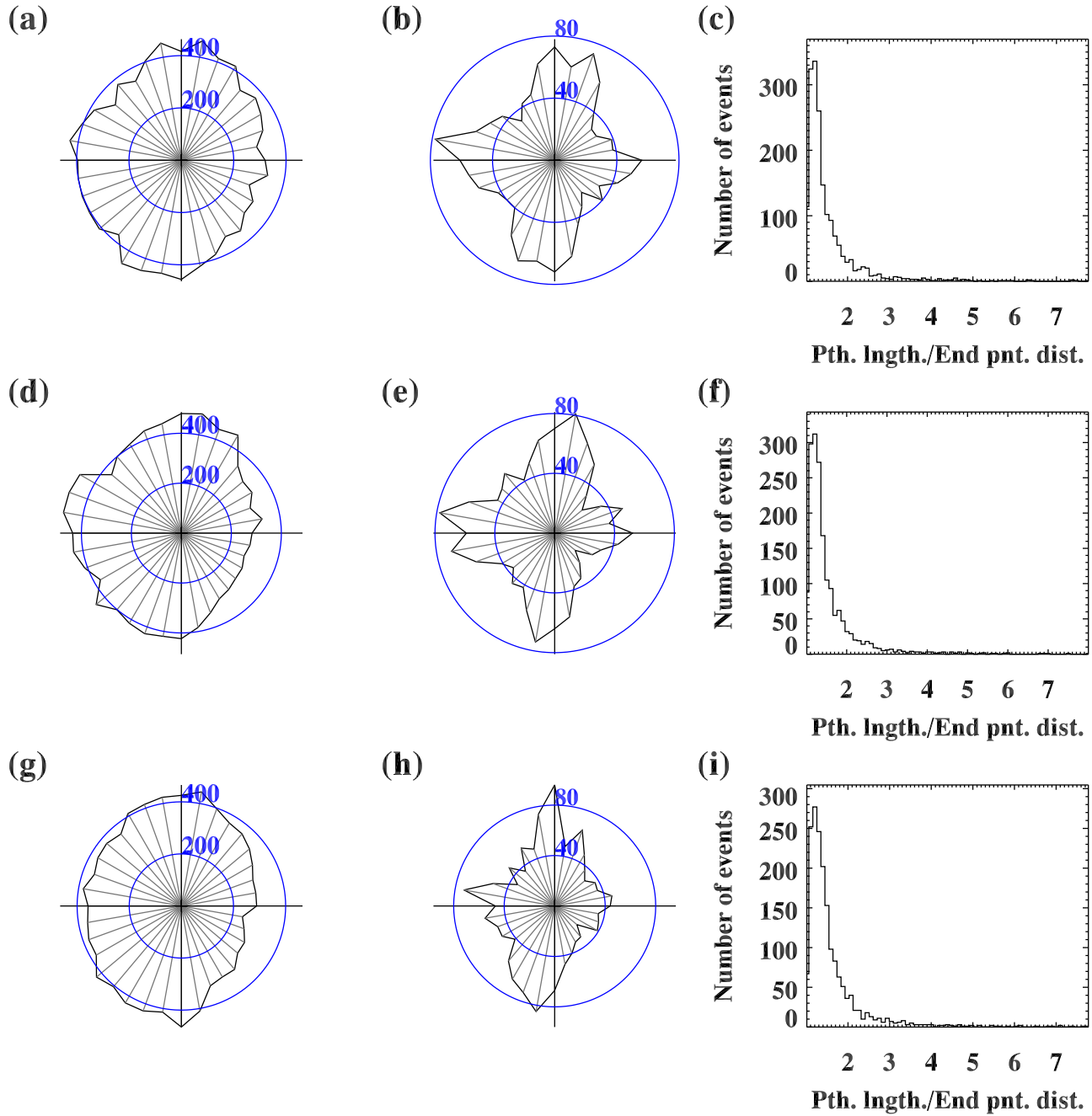}
    \caption{The top, middle, and bottom row of this figure correspond to channels 1400, 1330, and 2796 \AA\ respectively. (a), (d), and (g) show the distribution of a centroid's direction of motion between consecutive time steps during the lifetime of all detections. At each 10\de\ bin, the radial distance from the center shows the number of cases moving at that angle, with the labelled blue circles showing the number of cases per angular bin. The angles in this graph correspond to the image orientation (so that the graph $x$- and $y$-axis correspond to the image $x$- and $y$-axis). (b), (e), and (h) shows the distribution of the centroid overall direction of motion, defined as the angle between the start and end centroid of an event. (c), (f), and (i) show the distribution of the ratio between the total path length for an event (summed over all time steps) and the distance between the start and end centroid of that event. This ratio is 1 if the event moves along a straight line, and increases with increasing deviation from a straight line.}
    \label{motionstats}
\end{figure}

Figure \ref{motionstats} presents some statistical distributions gained from the detection's motion tracking. Between consecutive time steps over an event's duration, we calculate the angle between the moving spatial centroid of the event. Collecting this information over the lifetime of all events give the angular distributions shown in the left column for channels 1400, 1330, and 2796 \AA\ (top to bottom). For all three channels, the event is skewed so that there are fewer events moving in the positive $x$ direction compared to the negative $x$ direction. The distribution in the vertical $y$ direction is close to symmetrical. The asymmetry in the $x$ direction is unexpected, and shows that the motions are not purely random. The middle column shows the distribution of angles between the start and end centroid positions of an event, thus related to the overall direction of motion of an event. Distributions are similar across all three channels, show a clear preference for non-diagonal motions, and skewed at around 10\de\ to the image $x$ and $y$ directions. The asymmetry in the $x$ direction is present in these distributions. The 2796 \AA\ distribution (panel h) shows a considerable preference for motions in the $y$ direction compared to the $x$. 

For each event, we calculate the distance between the start and end centroid location, and calculate the total path length of the centroid position summed over all time steps. The ratio of the total path length to the start-end straight line distance gives a measure of an event's deviation from a straight line, and the distributions of this ratio are shown in the right column of figure \ref{motionstats}. The distributions show that most events move along lines that are close to straight (in the image plane) with the distribution peaking close to a ratio of 1.15 in all 3 channels, although large number of events follow curves that deviate considerably. The mean (median) ratio is close to 1.4 (1.6) for all 3 channels. We emphasise that these are the image plane motions, and that without an analysis of spectra the line of sight motion is unknown. 

\begin{figure}
    \centering
    \includegraphics[width=0.98\textwidth]{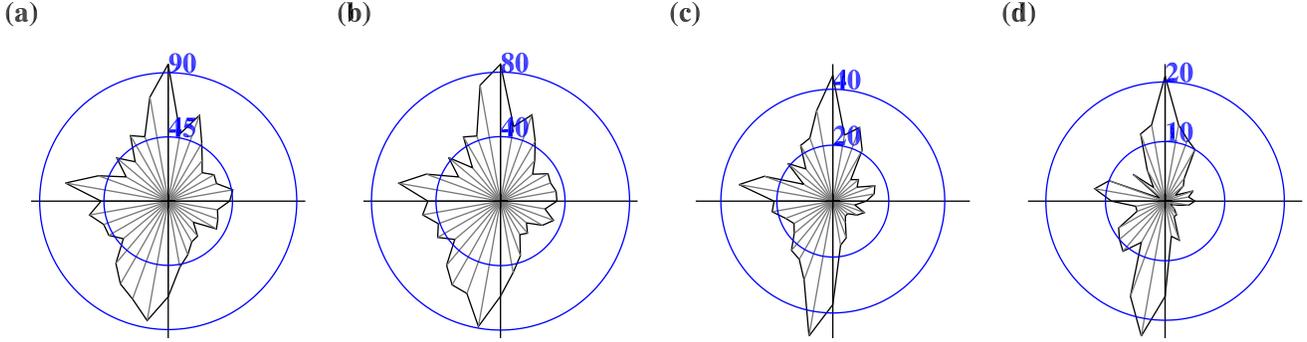}
    \caption{Results showing how the angular distributions change as we include only events that travel distances greater than a minimum threshold, set at (a) 0 (all events), (b) 2.5 pixels (1579 events), (c) 5 pixels (725 events), and (d) 7.5 pixels (213 events). These distributions are for the 2796\AA\ channel, and the angle is defined between the start and end centroid of events (so that panel (a) is the same as figure \ref{motionstats}h).}
    \label{motionstats2}
\end{figure}

Several points of interest arise from figure \ref{motionstats}. The distributions show that motions are not random. Within an event's lifetime, the motion is more likely to follow a straight line, or deviate only by small values from a straight line. There is a preference to move in the negative $x$ direction, and there is a strong directional preference for the overall direction of motion defined by the start and end points of an event. We investigated the cross-shaped angular distribution of the central column of figure \ref{motionstats} by filtering events according to the distance travelled during the event lifetime. As we discarded events that travelled only short distances, this cross became more pronounced. This result is shown for the 2796\AA\ channel in figure \ref{motionstats2}. When we limit events to those that travel a distance of at least 7.5 pixels, as shown in figure \ref{motionstats2}d, the vast majority of these events have an overall motion in directions close to the image vertical (solar north-south). These statistical characteristics are likely related to the orientation of the magnetic field in the relevant atmospheric layers, with direction of motions along the field perhaps related to pressure (gas and magnetic) gradients. When we select events that travel a greater distance, we may be selecting events associated with magnetic field structures of greater length, and these larger scale fields may tend to align in a particular direction over the field of view of the IRIS slitjaw image. Events that move over a smaller distance may be associated with the smaller-scale magnetic field loops that have a more even angular distribution. Validating this would require comparison with simulations, based on a magnetic field extrapolated from photospheric measurements, and is challenging considering the small spatial scales of the detections.

To gain meaningful results from the PoS motion information, future work demands a statistical approach. Further interpretation is uncertain without establishing whether there are certain common patterns over a large sample. In just this one dataset, we have thousands of 3-channel detections, thus an automated approach is required. Developing an automated routine that can record and group patterns of PoS motion across multiple channels will lead to valuable new diagnostics and would give: (i) robust confidence levels on the causal connections between multi-channel detections, (ii) classification into different types of events, and (iii) physical insight into the behaviour of different types.

\subsection{Comparison with coronal properties}
\begin{figure}
    \centering
    \includegraphics[width=0.7\textwidth]{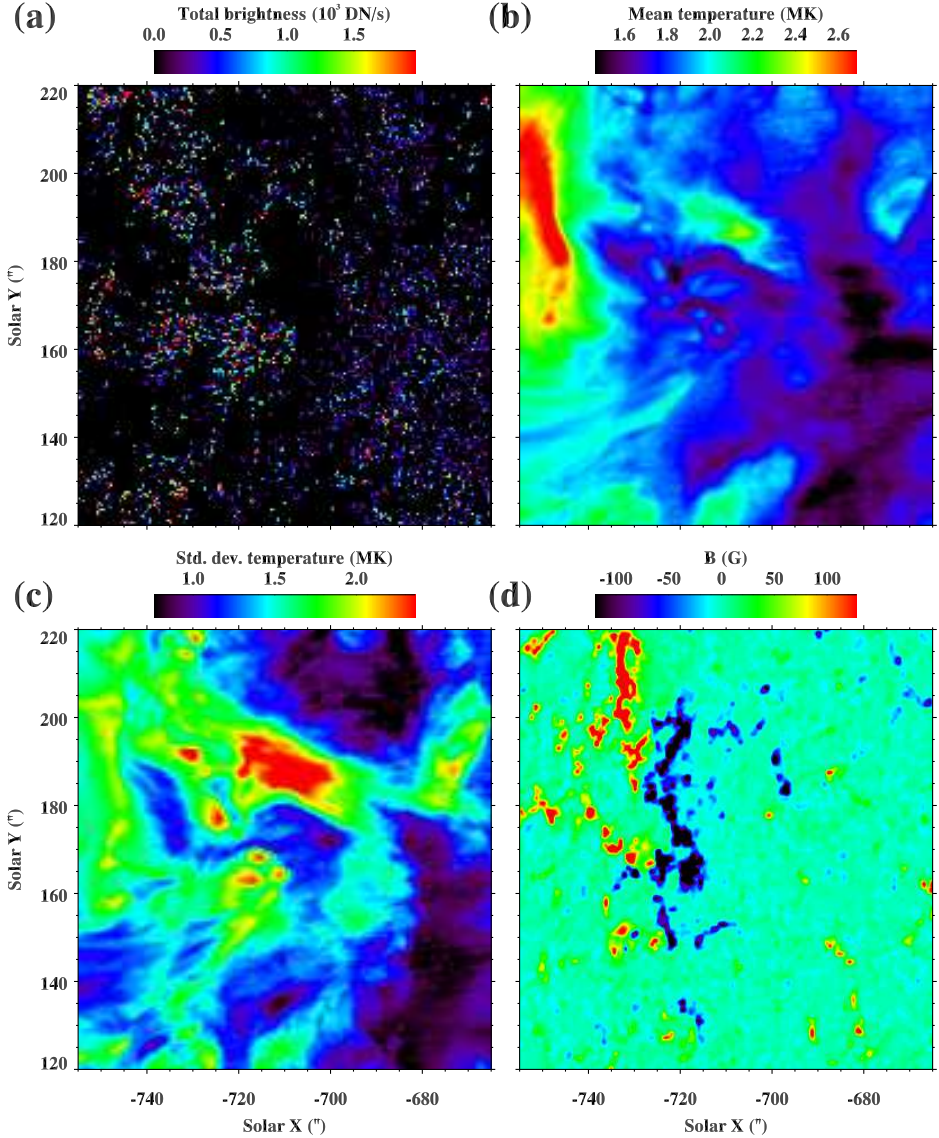}
    \caption{(a) The spatial distribution of cumulative total brightness of IRIS brightening events over the time of observation. (b) The mean temperature as derived from AIA/SDO observations. The temperature is a DEM-weighted mean (see text). (c) The standard deviation of temperature, DEM-weighted (see text). This is a measure of the multi-thermality of the DEM. (d) The integrated DEM, or EM, over the coronal temperatures accessible by AIA/SDO.}
    \label{aia}
\end{figure}

Figure \ref{aia}a shows the spatial distribution of brightening events detected over the whole time of observation. This shows the cumulative brightness, summed over all brightenings within each spatial bin. In order to collect meaningful statistics, we have summed over $5\times5$ IRIS pixels for this distribution. The distribution is interesting: there are broad regions where no brightenings have occurred, including a central region west of the active region, associated with the small coronal hole. Other regions void of events include areas bounding the small active region to the east, and a patch in the north-central field of view that also bounds the active region. The largest collection of brightenings is above the active region. We next compare this distribution to certain observable characteristics of the overlying corona and underlying photosphere.

Figure \ref{aia}b shows the mean coronal temperature gained from a differential emission measure (DEM) analysis of data from the Atmospheric Imaging Assembly \citep[AIA:][]{lemen2012} aboard the Solar Dynamics Observatory \citep[SDO:][]{pesnell2012}. The data of the EUV channels of AIA is provided by the IRIS Lockheed Martin Solar \&\ Astrophysics Laboratory database, conveniently cropped in space and time to match the IRIS data. For each channel, the mean image over the time of observation is calculated, and the DEM calculated using the Solar Iterative Temperature Emission Solver (SITES, \citet{morgan2019,pickering2019}). The mean temperature is calculated using a weighted mean over emission \citep{morgan2017}. The mean temperature distribution is as expected: higher temperatures overlie the active region to the east, and low temperatures overlie the coronal hole. There is no immediate correlation with the distribution of the IRIS brightenings, other than the active region is host to the most brightenings and the highest coronal temperatures. 

Figure \ref{aia}c shows the standard deviation of temperatures, again weighted by the DEM. This is the spread of temperatures around the mean, weighted by the DEM, and is a proxy for the multithermality of the corona. In this case, there is some correlation with the distribution of IRIS brightenings. In the region at $x=-750$\arcsec\ and $y=155$ to $180$\arcsec, a narrow line of high IRIS brightening activity, aligned approximately northwards, corresponds closely to a line of high multithermality in the AIA DEM. The region from $x=-730$ to $-710$\arcsec\ and $y=155$ to $170$\arcsec\ also shows good correspondence. The region of no IRIS events near the image center has a corresponding region of low multithermality, although the shape and extent of the region is different. Note we cannot expect an exact spatial correspondence since the corona is linked to the lower atmospheric layers by a non-radial magnetic field, and the region of study is not near disk center (so that lines of sight through the corona are not close to the solar normal). These results suggest that whilst the mean temperature of the corona is not strongly linked to the distribution of small-scale brightening activity in the lower atmosphere, the spread of the DEM across temperatures tends to be greater above regions of high brightening activity. There may be a link, therefore, between brightening events in the lower atmosphere and multithermality in the corona. This needs to be confirmed with a larger data sample and a more detailed analysis. 

Figure \ref{aia}d shows the photospheric magnetic field as derived from the Helioseismic Magnetic Imager \citep[HMI:][]{scherrer2012,schou2012}/SDO data. Below the eastern active region are patches of high field strength, with polarities distributed in an approximately north-south alignment either side of the $x=-725$\arcsec\ line. It seems as if this region of high field (and high field gradients) does correspond to regions of high IRIS activity. The patches of strong negative field just north of the central coronal hole also seem to underlie a region of enhanced IRIS activity.

Although no definitive or clear conclusions can be drawn, this section shows that there may be a tentative connection between the spatial distribution of brightening activities in IRIS with the properties of higher atmospheric layers. A proper treatment would require a mapping of the magnetic field between the different layers, in order to compare connected regions, and a statistical analysis of many datasets. Whilst a spatial correspondence between IRIS activity and multithermality in one observation does not prove a causal connection, these results show the need for further study.

%==============================
% Conclusions
%==============================

\section{Conclusions}\label{sec:study_3_conclusions}

We present the results of our general filter/threshold method for detecting small-scale brightenings on real IRIS AR data. 
Having applied the method to the 1400 \AA\ data channel, new \threshlo\ and \threshhi\ values are set for the 1330 \AA\ and 2796 \AA\ channels based on detecting the same number of events as those in the 1400 \AA\ channel plus 10\%. These over-detected results are then matched and trimmed down to match the number of 1400 \AA\ results using the minimum spatio-temporal separation between each event. Events in different channels are paired using a stringent statistical criterion. This leaves 2377 multi-wavelength detections in a $90\arcsec\times100\arcsec$ FOV over $\sim34.5$ minutes. $\approx1800$ of these remain unfragmented or fragment only once over their lifetime, yielding an event density range of 9.60$\times10^{-5}$ to 9.85$\times10^{-5}$ over all wavelength detections. 

A correlation between the $N_{frag}$ parameter, average area and duration is evident, while no correlation appears to exist between $N_{frag}$, average speed and maximum brightness. A weak relationship (with a LAD-fitted gradient of 0.22) is present between $N_{frag}$ and total brightness. Thus brighter, larger, and longer-duration events are more likely to fragment to multiple parts over their lifetime. Difference histograms demonstrate that the events in three channels have similar area, speed, and duration. 1400 \AA\ events' maximum and total brightness are typically larger than those of 1330 \AA\ and 2796 \AA\ - this is largely due to the difference in effective area of each wavelength channel intrinsic to the IRIS instrument. Mean values of area, speed and duration are determined as 0.386-0.414 arcsec$^{2}$, 7.22-7.36\kms\ and 161-181s, respectively, with the ranges arising from the variance from each channel.

The total brightness, maximum brightness and average area distributions of these events obey a power-law while their average speed and duration distributions do not. These gradients of the power laws lie within $2.78<\alpha<3.71$ for total brightness, $3.84<\alpha<4.70$ for maximum brightness and $4.31<\alpha<5.70$ for area.  

Every event is moving, with a minimum speed of 1.4-1.9\kms\ and a median of $\sim$7\kms. Examples of the PoS motion paths of events are interesting, and show a powerful new diagnostic that can be applied statistically to a large number of events. Many cases seem to show the two hotter channels separated in position and/or trajectory from the cooler 2796 \AA\ event, with the cooler event appearing at a slightly later time. Explanations for this is an underlying impulsive event followed by rapid cooling, or events occurring at the TR and a subsequent propagating disturbance to lower and cooler atmospheric layers.

We briefly explore the spatial distribution of the bright points, and find a non-uniform distribution, with many broad regions devoid of activity. Our comparison shows that there seems to be some correspondence between the distribution of IRIS events and the multithermality of the overlying corona although we cannot show a direct causal connection. This aspect requires further study.

Having applied this filtering and thresholding method to a multi-wavelength IRIS data set, we plan to apply it to several more sit-and-state IRIS data, including those that are more energetic and active. However, suitable multi-wavelength sit-and-stare IRIS observations are infrequent which may hinder any comprehensive multi-wavelength analyses. Future studies will also implement co-aligned data from other EUV imaging instruments and/or ground-based photospheric observations to determine whether these events posses H$\alpha$ or other wavelength counterparts. A further investigation in to the nature of the events' PoS motions and their $N_{frag}$ parameter is also of great interest. The wealth of information provided by the method provides scope for more detailed future analysis, including categorization of events into different types, and placing constraints on the events formation mechanism. For example, can the information provided by the method identify whether formation mechanisms are based on small-scale reconnections or wave-induced shocks? This will involve developing parameters that separate the events into clusters based on e.g. shape, size, duration, intensity, motion, and tendency to fragment.
The method and software can easily be adapted for use on other types of imaging observations, given appropriate values for a small number of parameters, namely: filter bandpass frequency limits, intensity thresholds, minimum volume, and minimum duration. For example, we envisage running the method on detecting moving bright dots in coronal EUV images as done by \cite{Alpert_2016} for Hi-C, or for sunspot bright dots or penumbral jets \citep{Katsukawa_2007, Tiwari_2016, Tiwari_2018, Drews_2017}.

\acknowledgments
We thank an anonymous referee for comments that greatly improved this work. 
We acknowledge (1) STFC grant ST/S000518/1 to Aberystwyth University which made this work possible; (2) STFC PhD studentship ST/S505225/1 to Aberystwyth University; and (3) a Coleg Cymraeg Cenedlaethol studentship award to Aberystwyth University.
IRIS is a NASA small explorer mission developed and operated by LMSAL with mission operations executed at the NASA Ames Research center and major contributions to downlink communications funded by ESA and the Norwegian Space Centre. 
We acknowledge Super Computing Wales for the provision of excellent computing facilities and support.

\vspace{5mm}
%\facilities{}
\bibliography{study_3}
\bibliographystyle{aasjournal}

\end{document}